\documentclass[aps,prb,
preprint,
10pt,
showpacs,
floatfix,
]{revtex4-1}
\usepackage{bm}
\usepackage{amsmath}    % For {align}
\DeclareMathOperator{\sech}{sech}

\usepackage{amssymb}    % Maybe unnecessary... \hslash
\usepackage{graphicx}   % For the figures
\usepackage[hang]{subfigure}
\newcommand{\jun}{junction }
\newcommand{\juns}{junctions }
\newcommand{\Jos}{Josephson }
\newcommand{\elli}{elliptic }

\begin{document}
\title[R.Monaco]{Engineering Double-Well Potentials \\with Variable-Width Annular Josephson Tunnel Junctions}
%\thanks{To be submitted to PRB}
\author{Roberto Monaco}
\email[Author's e-mail address:]{ r.monaco@isasi.cnr.it and roberto.monaco@cnr.it}
%\email[Corresponding author e-mail address:]{ r.monaco@isasi.cnr.it}
%%\author{Roberto Monaco$^{\dag}$, Carmine Granata, and Antonio Vettoliere}
%\address{CNR-ISASI, Institute of Applied Sciences and Intelligent Systems ''E. Caianello'', Comprensorio Olivetti, 80078 Pozzuoli, Italy}
%$\qquad \qquad \qquad ^{\dag}$ \textit{corresponding author}
%\ead{roberto.monaco@cnr.it}
%%http://publish.aps.org/authors/length-guide

\affiliation{CNR-ISASI, Institute of Applied Sciences and Intelligent Systems ''E. Caianello'', Comprensorio Olivetti, 80078 Pozzuoli, Italy}

\pacs{85.25.Cp,03.67.Lx,05.45.Yv,03.65.Ge}
\date{\today}

\begin{abstract}
\vskip 4pt
\begin{large}
%trusty, trustworthy  a potential structure which spatially confines the electron. 
Long Josephson tunnel junction are non-linear transmission lines that allow propagation of current vortices (fluxons) and electromagnetic waves and are used in various applications within superconductive electronics. Recently, the Josephson vortex has been proposed as a new superconducting qubit. We describe a simple method to create a double-well potential for an individual fluxon trapped in a long elliptic annular Josephson tunnel junction characterized by an intrinsic non-uniform width. The distance between the potential wells and the height of the inter-well potential barrier are controlled by the strength of an in-plane magnetic field. The manipulation of the vortex states can be achieved by applying a proper current ramp across the junction. The read-out of the state is accomplished by measuring the vortex depinning current in a small magnetic field. An accurate one-dimensional sine-Gordon model for this strongly non-linear system is presented, from which we calculate the position-dependent fluxon rest-mass, its Hamiltonian density and the corresponding trajectories in the phase space. We examine the dependence of the potential properties on the annulus eccentricity and its electrical parameters and address the requirements for observing quantum-mechanical effects, as discrete energy levels and tunneling, in this two-state system.
\end{large}
%A magnetic field applied in the junction plane gives rise to a tunable non-sinusoidal periodic potentials. Transverse magnetic field.
%Alternative classical explanations have been suggested to explain 
\end{abstract}
\maketitle
%
%
%%%%%%
\vskip -8pt
\listoffigures
\vskip -18pt
\tableofcontents
\vskip -8pt
\newpage
%%%%%
\section{Introduction}

The quantum tunneling in a double-well potential, in spite of being almost as old as quantum mechanics \cite{hund27,morse,dennison32}, is still one a very active research area. For example, the tunneling dynamics appeared in the mean-fields dynamics of Bose-Einstein condensates \cite{Ligner,Theocaris}, the development of ion trap technology \cite{Retzker}, the ultra-cold trapped atoms theory \cite{Dounas-Frazer} and optical systems applications \cite{Grossman,DellaValle}. The double-well potential is currently considered for possible detection of macroscopic quantum effects and the realization of viable quantum bits (qubits, i.e., two-state quantum-mechanical systems) for information processing. Recent successes with various types of superconducting qubits have enhanced the feasibility of implementing quantum computing operations, such as the factorization of an integer number into its constituent primes, with Josephson devices \cite{Lucero}. Rabi oscillations, namely the oscillations in the population of the first excited level as a function of the applied microwave power, which are a preliminary requirement of quantum computing, have been reported in charge \cite{Makhlin,You}, phase \cite{Yu}, and flux qubits \cite{Mooij}. The operation of these systems is based on quantum coherence of the charge state, the Josephson phase difference, or the magnetic-flux state, respectively, in circuits made of short \Jos tunnel junctions (JTJs). Long JTJs with quantized vortices of supercurrent, also called \textit{fluxons}, which are particle-like collective nonlinear excitations of the phase difference, have also been proposed to observe macroscopic quantum tunneling \cite{kato96,malomed97} and to implement Josephson vortex qubits \cite{wallraff00,shaju05} as they are well decoupled from other electromagnetic excitations. At milli-Kelvin temperatures, fluxons display macroscopic quantum properties and exhibit quantum tunneling \cite{wallraff03}. Furthermore, preparation and readout of the vortex state have been reported in annular JTLs where a double well potential was experimentally realized. Up to now, three types of prototypes have been investigated for the realization of vortex qubits with long and narrow (planar) JTJs, often also called Josephson transmission lines (JTLs). The first is the heart-shaped JTJ \cite{kemp02} in which the effective potential experienced by the vortex along the length of the junction is formed by the interaction of the vortex magnetic moment with the uniform external magnetic field. The second method uses a linear JTL with local magnetic fields generated by control current injectors to create a desired vortex potential \cite{fistulPRB03}. One more prototype is based on the interaction of the vortex with the potential created by a localized microshort \cite{shaju04,shaju05}, a thin spot in the dielectric barrier with enhanced Josephson current implemented via a section of insulating barrier that is locally wider in the junction plane \cite{kemp10}; here, a double-well potential for the vortex is created by the competition between the repulsion at the microshort and pinning by an in-plane magnetic field. The attractive potential of microresistors \cite{kim11} (i.e., a narrow regions with reduced $J_c$) that can be implemented by localized width reductions has never been realized. All these prototypes are modeled by a perturbed sine-Gordon equation that determines the spatial and temporal behavior of the \Jos phase; however, severe approximations are required to take into account the discontinuities in the junction parameters and/or in the component of the magnetic field normal to junction perimeter. In order to achieve a large designing reliability, it is, therefore, advantageous to look for geometrical configurations where all quantities are smoothly distributed along the JTL length, so guaranteeing an accurate modeling. 

%In the relativistic theory, it is $E=E_0+KE=\gamma m$ with $\gamma=1/\sqrt{1-u^2}$, $E_0=mc^2$ is the rest energy and $KE=(\gamma-1)mc^2$ is the kinetic energy.

\begin{figure}[tb]
\centering
\subfigure[ ]{\includegraphics[height=4cm,width=7cm]{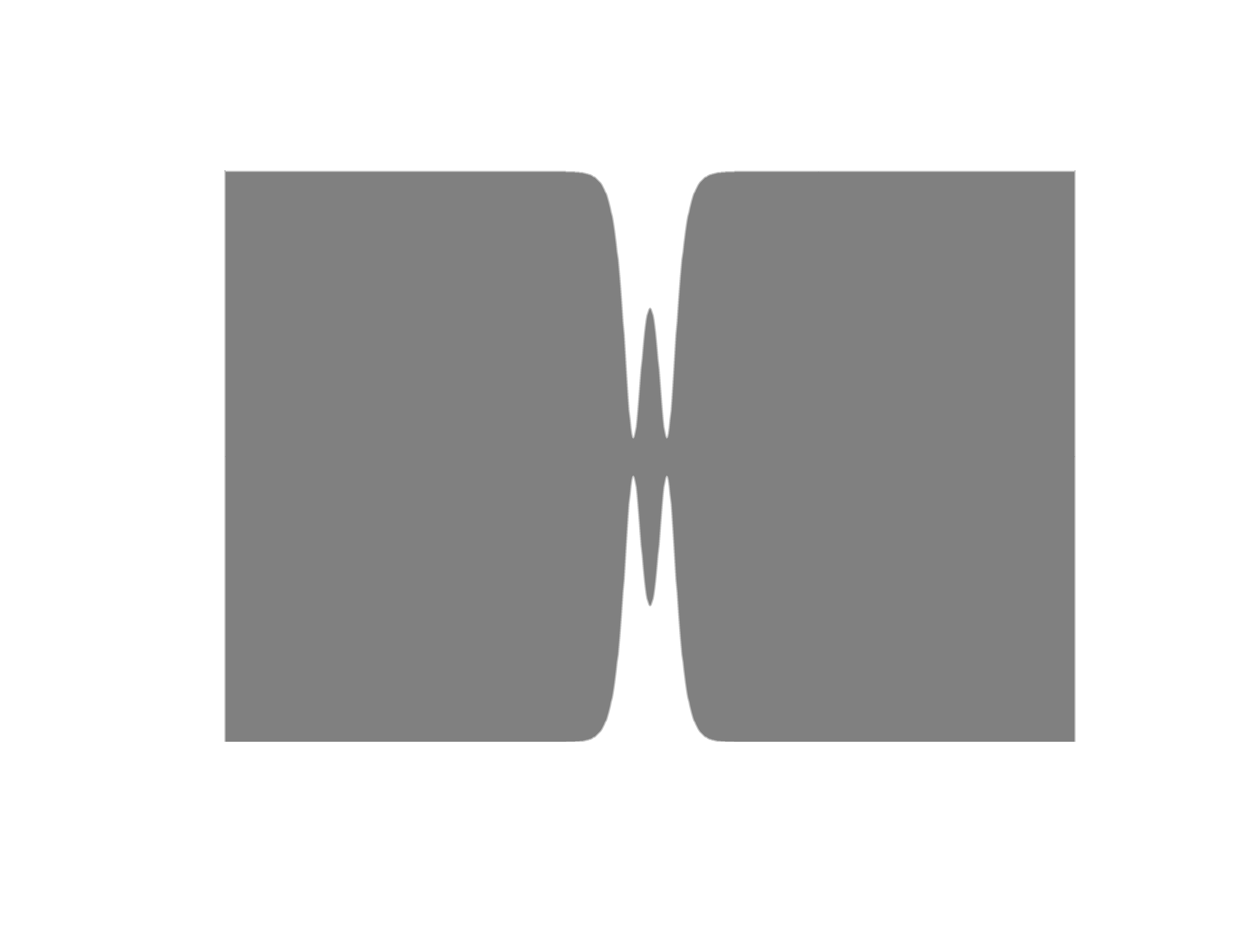}}
\subfigure[ ]{\includegraphics[height=4cm,width=7cm]{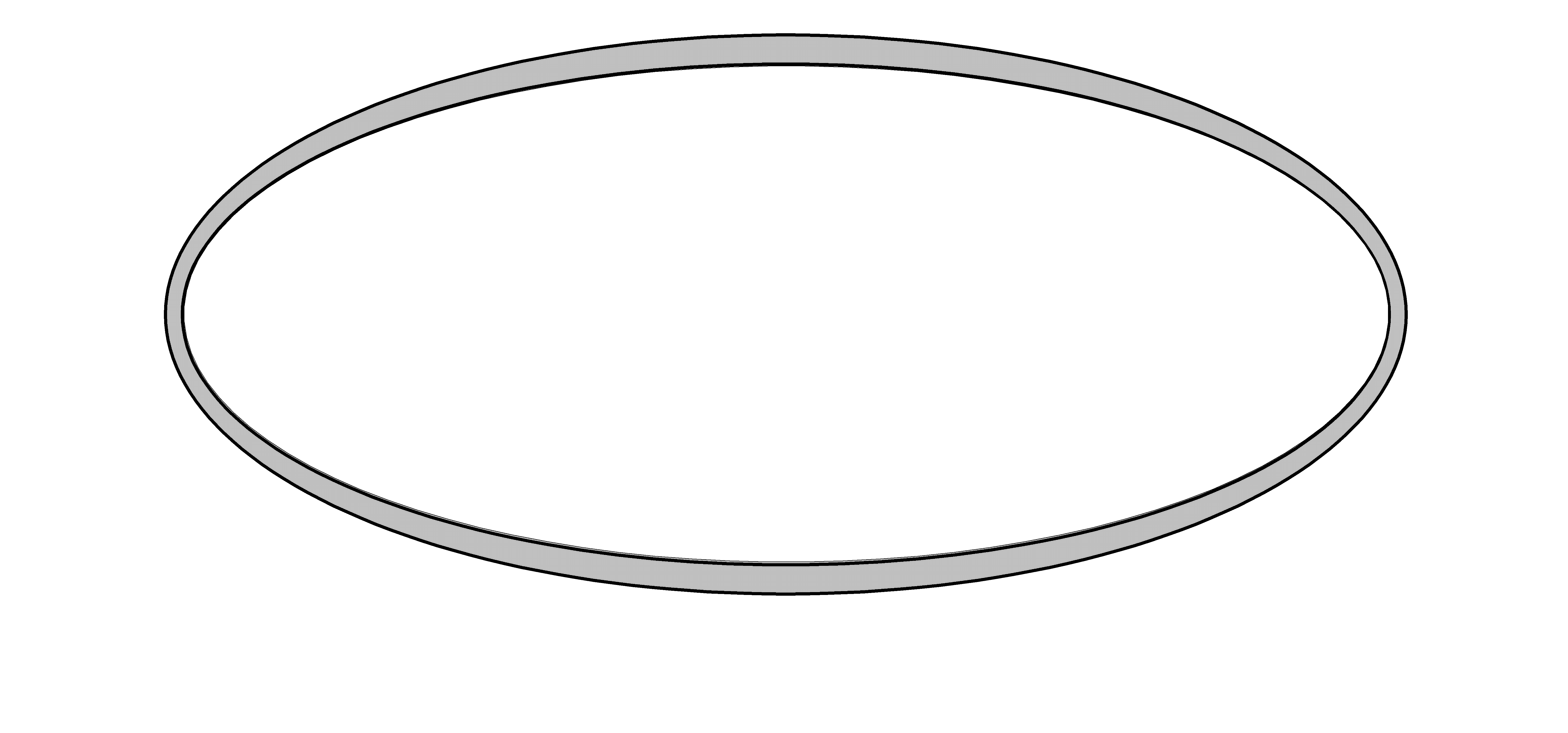}}
\caption{Tunneling area of variable-width long planar \Jos tunnel junctions (the junction electrodes are not shown): (a) sketch (not to scale) of a variable-width linear JTL (the direction of fluxon propagation is horizontal); (b) annular JTL delimited by two confocal ellipses given in Eqs.(\ref{inner}) and (\ref{outer}). The annulus width is smallest at the equatorial points and largest at the poles.}
\label{JTL}
\end{figure}

It has been addressed long ago \cite{scott} that the energy of a fluxon traveling with constant speed on a lossless infinite JTL is $\hat{E}=8/\sqrt {1-\hat{u}^2}$, where $\hat{u}$ is the fluxon speed normalized to the Swihart velocity \cite{Swihart}, $\bar{c}$, which is the characteristic velocity of electromagnetic waves in JTJs. For non-relativistic speeds the fluxon energy becomes $\hat{E}=8(1+\hat{u}^2/2)$. Throughout the paper we use circumflex accents to denote both normalized quantities and unit vectors. $\hat{E}$ is normalized to the characteristic energy, $\mathcal{E}=\Phi_0 J_c \lambda_J W/2\pi$, where $\Phi_0$ is the magnetic flux quantum, $J_c$ is the maximum \Jos current density and $\lambda_J$, called the \Jos penetration length of the junction, gives a measure of the distance over which significant spatial variations of the \Jos phase occur, typically, of the order of several micrometers. Therefore, the fluxon rest mass is $m=8\mathcal{E}/\bar{c}^2\propto W$. $W << \lambda_J$ is the JTL width that is assumed to be constant; typically, for submicrometer-width JTLs, $m$ is about thousand times smaller than the electron rest mass \cite{kato96}. Then $8\mathcal{E}$ is the energy of a JTL containing one individual static fluxon. In a consistent manner, Nappi and Pagano \cite{nappipagano} demonstrated that the potential energy of a (non-relativistic) fluxon on a variable-width JTL is proportional to the local width, $\hat{U}(s)=8\hat{w}(s)$, where $\hat{w}$ is some normalized width and $s$ is a curvilinear coordinate, allowing the JTL to be curved. It follows that a large variety of intrinsic spatially dependent fluxon potentials can be engineered by means of JTLs having a non-uniform width, even in the absence of an external magnetic field. An example is sketched in Figure~\ref{JTL}(a) where the width of a linear JTL is tailored to implement the so-called \textit{double P\"{o}schl-Teller potential}, $U^*(s)=U_{\infty}-U_{min} [\sech^2(s-\bar{s})/a +\sech^2(s+\bar{s})/a]$, with two symmetric minima, $U_{\infty}-U_{min}$, in $\pm \bar{s}$ (the single-well P\"{o}schl-Teller potential is one of the few exactly solvable potentials in quantum mechanics \cite{landau}). A classical particle traveling in a position-dependent potential also has a position-dependent kinetic energy. As we will see, the case of variable-width JTLs is made more interesting by the fact that the traveling fluxon also has a position-dependent inertial mass. If we cut the JTL in Figure~\ref{JTL}(a) along the horizontal line of symmetry, the potential $U^*$ is preserved. A JTL can be bent and its extremities can be jointed to form a doubly-connected or annular JTL; then the boundary conditions of the open simply-connected configuration are replaced by periodic conditions. A unique property of not simply-connected \juns is due to the fluxoid quantization in the superconducting loop formed by either the top or the bottom electrodes of the tunnel junction. One or more fluxons may be trapped in the \jun during the normal-superconducting transition. Once trapped the fluxons can never disappear and only fluxon-antifluxon pairs can be nucleated. In the presence of an in-plane magnetic field, ring-shaped JTLs were recognized to be ideal devices to investigate both the statics and the dynamics of sine-Gordon solitons in a spatially periodic potential \cite{gronbech, ustinov,PRB98,wallraff03}. The annulus shape does not need to be circular: also heart-shaped junctions can be classified as annular JTLs; the only requirement for an accurate modeling is that the curvature radius of the annulus is everywhere larger than $\lambda_J$. 
%$\mathcal{E}$ is also the Josephson coupling energy of a small Josephson junction of the area $\lambda_J W$.

Very recently, the static and dynamic properties were investigated for constant \cite{SUST15,JLTP16a} and variable \cite{JLTP16b} width elliptic annular JTLs in the presence of an in-plane magnetic field of arbitrary orientations. In both cases, the fluxon motion is characterized by a strong radial inward acceleration where the curvature radius is smallest. However, in variable-width elliptic annuli the fluxon acceleration results not only from the curvilinear motion, but also from a periodic variation of the fluxon tangential speed (even in the absence of an external magnetic field). The numerical analysis showed that both the fluxon statics and dynamics are strongly affected by the non-uniform width of the annulus. The variable-width elliptic annular JTLs were named Confocal Annular Josephson Tunnel Junctions (CAJTJs) since they are delimited by two ellipses having the same foci \cite{JLTP16b}. As depicted in Figure~\ref{JTL}(b), in this configuration the width variation is smoothly distributed along the JTL perimeter which is advantageous for the realization of an intrinsic robust symmetric double-well potentials. The paper is organized as follows. In the rest of this Section we state the problem by describing the geometrical properties of a CAJTJ and introduce the mathematical notations and identities used throughout this paper. In Sec. II, we present the modeling framework of our study, which is based on a modified and perturbed sine-Gordon equation that, although not integrable, can be derived by an analytic Lagrangian density corresponding to a conservative Hamiltonian: we shows that in a CAJTJ the fluxon has a position-dependent inertial mass and calculate its classical trajectories in the phase space; later on we discuss the conditions under which CAJTJs enter the quantum regime. In Sec. III we will extend the analysis to take into account the perturbative effects of an external magnetic field and of a bias current; in particular, we demonstrate that: i) a small field creates a double-minima potential whose inter-well barrier shrinks with increasing field strength and ii) a small current tilts the fluxon potential making one of the equilibrium positions metastable. After that, we will examine the fluxon tunneling time and probability in the quantum limit. In Sec. IV, we present numerical simulations concerning the fluxon static and dynamic properties and describe a protocol to reliably determine and prepare the vortex state. The conclusions are drawn in Section V.

\subsection{Confocal Annular \Jos Tunnel Junctions (CAJTJs)}
% \url{http://demonstrations.wolfram.com/ConstantCoordinateCurvesForEllipticCoordinates/}

\noindent We first introduce the (planar) confocal elliptic coordinates $(\nu,\tau)$, with $\nu\geq0$ and $\tau\in[-\pi,\pi]$, such that, any point $(x,y)$ in the $X$-$Y$ plane is uniquely expressed as $(c\cosh\nu\sin\tau, c\sinh\nu\cos\tau)$ with $\nu\geq0$ and $\tau\in[-\pi,\pi]$ for a positive $c$ value. In the limit $c\to0$, the elliptic coordinates $(\nu,\tau)$ reduce to polar coordinates $(r,\theta)$; the correspondence is given by $\tau\to \theta$ and $c\cosh\nu\to r$ (note that $\nu$ itself becomes infinite as $c\to0$). In elliptic coordinates the elementary distance is $ds= \sqrt{dx^2+dy^2} =f(\nu,\tau)\sqrt{d\nu^2+ d\tau^2}$, where $f(\nu,\tau)= c\,q(\nu,\tau)$ is the so-called scale factor with $q^2(\nu,\tau)\equiv \sinh^2\nu \sin^2\tau+\cosh^2 \nu \cos^2 \tau= \sinh^2\nu+ \cos^2\tau=\cosh^2\nu - \sin^2\tau=(\cosh2\nu + \cos2\tau)/2$. Furthermore, the elementary surface element is $dS=dxdy=f^2d\nu d\tau$. Any vector ${\bf H}$ applied at a point $(\nu,\tau)$ can be decomposed in its normal and tangential components, respectively, $H_\nu={\bf {H}}\cdot{\bf \hat{N}}$ and $H_\tau={\bf {H}}\cdot{\bf \hat{T}}$, were:
%[[Furthermore, it is $\cosh\,2\nu=(\coth^2\nu-\tanh^2\nu)/(\coth\nu-\tanh\nu)^2$ and $2\coth2\nu=\coth\nu + \tanh\nu$. It is also $q^2(\nu,\tau)= \cos(\tau + \hat{\imath}\nu) \cos(\tau- \hat{\imath}\nu)= \cosh(\nu+ \hat{\imath}\tau)\cosh(\nu-\hat{\imath}\tau)$.]]

\begin{subequations}
\begin{eqnarray}
{\bf{\hat{N}}}\equiv&[\frac{\sinh\nu\sin\tau}{q(\nu,\tau)},\frac{\cosh\nu \cos\tau}{q(\nu,\tau)}], \label{nu} \\
{\bf \hat{T}}\equiv&\,[\frac{\cosh\nu \cos\tau}{q(\nu,\tau)}, -\frac{\sinh\nu\sin\tau}{q(\nu,\tau)}], \label{tau}
\end{eqnarray}
\end{subequations}
 
\noindent are, respectively, the (outward) normal and (clockwise) tangent unit vectors to the ellipse passing at the point $(\nu,\tau)$. 
 
Let us now consider two ellipses centered in the origin of a Cartesian coordinate system $X$-$Y$ and whose principal diameters are parallel to the $X$ and $Y$ axes. If the ellipses are confocal and the foci $(\pm c,0)$ lie on the $X$-axis, the parametric equations of the inner and outer ellipses are, respectively,: 
\vskip -8pt
\begin{equation}
\label{inner}
\begin{cases}
x_i(\tau)=a_i\sin\tau=c\cosh\nu_i\sin\tau; \\
y_i(\tau)=b_i\cos\tau=c\sinh\nu_i\cos\tau; \end{cases} 
\end{equation}
\vskip -8pt
\noindent and
\vskip -8pt
\begin{equation}
\label{outer}
\begin{cases}
x_o(\tau)=a_o\sin\tau=c\cosh\nu_o\sin\tau; \\
y_o(\tau)=b_i\cos\tau=c\sinh\nu_o\cos\tau, \end{cases} 
\end{equation}
\vskip -4pt
\noindent where $\nu_o>\nu_i$; $2a_{i,o}$ and $2b_{i,o}$ are, respectively, the mayor and minor ellipses diameters and $\tau$ is a parameter measured clockwise from the positive $Y$-axis. By definition, the area of a CAJTJ is the annulus delimited by the ellipses in Eqs.(\ref{inner}) and (\ref{outer}); the annulus width $\Delta w(\tau)$ is a $\pi$-periodic function of $\tau$:
\vskip -8pt
$$\Delta w(\tau)\!=\!c\sqrt{(x_o-x_i)^2+(y_o-y_i)^2}\!=\! c\sqrt{(\cosh\!\nu_o-\cosh\!\nu_i)^2\sin^2\!\tau +(\sinh\!\nu_o-\sinh\!\nu_i)^2\cos^2\!\tau}.$$

\noindent If $\Delta\nu\equiv \nu_o-\nu_i<<1$, the expression of the width reduces to:
\vskip -8pt
\begin{equation}
%\nonumber
\Delta w(\tau)=c\mathcal{Q}(\tau)\,\Delta\nu,
\label{width}
\end{equation}
\vskip -5pt

\noindent where $\mathcal{Q}(\tau) \equiv q(\bar{\nu},\tau)$ with $\bar{\nu}= (\nu_o+\nu_i) /2$. The maximum width value is $\Delta w_{max}=c\cosh\bar{\nu} \Delta\nu$ at the ellipse poles, $\tau=m\pi$ ($m$ integer), while $\Delta w_{min}=c\sinh\bar{\nu} \Delta\nu$ is the minimum value achieved at the equatorial points, $\tau=m\pi\pm \pi/2$. The width relative variation, $(\Delta w_{max}-\Delta w_{min})/\Delta w_{min} =\coth \bar{\nu}-1$, diverges as $\bar{\nu} \to 0 $. We define the axes (mean) ratio as $\rho\equiv \tanh\bar{\nu}$ and the annulus (mean) eccentricity as $e^2 \equiv 1-\rho^2=\sech^2\bar{\nu}$. When the eccentricity vanishes, the CAJTJ reduces to the well-known circular annular \Jos tunnel junction ideal for experimental tests of the perturbation models developed to take into account the dissipative effects in the propagation with no collisions of sine-Gordon kinks \cite{davidson, dueholm,hue}. Throughout the paper, by way of example, we will often select the moderate-eccentricity value $\bar{\nu}\approx 0.55$, for which $\coth \bar{\nu}=2$, such that the largest CAJTJ width is twice its smallest one; in fact, $\Delta w_{max}/\Delta w_{min}=\coth \bar{\nu}=1/\rho$.

%\noindent [[ If $b>a$, then the foci lie on the $Y$-axis and $c$ is an imaginary number, while, being $\rho>1$, $\bar{\nu}=\arctan\!\textrm{h}\,\rho= \arctan\!\textrm{h}\,\rho^{-1} -\hat{\imath} \pi/2$ is a complex number. In this case $x=a\sin\tau= c\sinh\nu\sin\tau$ and $y=b\cos\tau= c\cosh\nu\cos\tau$  with $\nu\geq0$. Being $\cosh^2 \nu-\sinh^2 \nu=1$, the foci ordinates are $c=\pm\sqrt{b^2-a^2}=\pm a\sqrt{\rho^2-1}$. Throughout the paper we will carry out the analysis assuming $a>b$; however, all the derived expressions will still be real when $a<b$, provided that $c$ is replaced by its imaginary counterpart, $\hat{\imath}\,c$. Some useful properties of the hyperbolic functions are: $\sinh(\nu - \hat{\imath}\pi/2)= -\hat{\imath}\cosh\nu$, $\cos(\nu - \hat{\imath}\pi/2)= -\hat{\imath}\sinh\nu$, $\sinh (2\nu - \hat{\imath}\pi)= -\sinh\nu$, and $\cos(2\nu - \hat{\imath}\pi)= -\cosh\nu$.]]

\section{Theory of one-dimensional CAJTJs} 

In the small width approximation, $\Delta w_{max}<< \lambda_J$, the \Jos phase does not depends on $\nu$ and the system becomes one-dimensional. Furthermore, the scale factor becomes $f(\bar{\nu},\tau)=c\, \mathcal{Q}(\tau)$ and the length of an elementary annulus arc is $ds=c \mathcal{Q}(\tau) d\tau$. Therefore, we introduce the non-linear curvilinear coordinate $s(\tau)=c \int_{0}^{\tau} \mathcal{Q}(\tau')d\tau'=c \cosh\bar{\nu}\,\texttt{E}(\tau,e^2)$, where $\texttt{E}(\tau,e^2)$ is the \textit{incomplete} elliptic integral of the second kind of modulus $e^2\leq1$. Accordingly, $s(\tau)$ increases by one (mean) perimeter, $L=\oint ds= 4c \cosh\bar{\nu}\,\texttt{E}(e^2)$, as $\tau$ changes by $2\pi$ (for a thin circular ring with mean radius $\bar{r}$, it would be $s(\tau)=s(\theta)=\bar{r} \theta$). $\texttt{E}(e^2)\equiv \texttt{E}(\pi/2,e^2)$ is the {\it complete} elliptic integrals of the second kind of argument $e^2$. Unfortunately, the \elli integrals of the second kind are not invertible in terms of single-valued functions (at variance with the \elli integrals of the first kind). In passing, we note that the elementary surface of the confocal annulus is $dS=c^2 \mathcal{Q}^2 \Delta\nu d\tau$, so that its area is $\Delta S=\pi c^2 \cosh2\bar{\nu}\Delta\nu$; furthermore, its perimeter can also be expressed as $L=2\pi \overline{\Delta w}/ \Delta\nu$, where $\overline{\Delta w}$ is the average width.

\noindent It has been recently derived that the radially independent \Jos phase, $\phi(\tau,\hat{t})$, of a CAJTJ in the presence of a spatially homogeneous in-plane magnetic field ${\bf H}$ of arbitrary orientation, $\bar{\theta}$, relative to the $Y$-axis, obeys a modified and perturbed sine-Gordon equation with a space dependent effective Josephson penetration length inversely proportional to the local junction width \cite{JLTP16b}:
\vskip -8pt
\begin{equation}
 \left[\frac{\lambda_J}{c\,\mathcal{Q}(\tau)}\right]^2 \left(1+\beta\frac{\partial}{\partial \hat{t}}\right) \phi_{\tau\tau} - \phi_{\hat{t}\hat{t}}-\sin \phi =\alpha \phi_{\hat{t}} - \gamma(\tau) + F_h(\tau),
\label{psge}
\end{equation}

\noindent where $\hat{t}$ is the time normalized to the inverse of the so-called (maximum) plasma frequency, $\omega_p^{-1}=\sqrt{\Phi_0 c_s/2\pi J_c}$ (with $c_s$ the specific junction capacitance) and the critical current density, $J_c$, was assumed to be constant. Here and in the following, the subscripts on $\phi$ are a shorthand for derivative with respect to the corresponding variable. Furthermore, $\gamma(\tau)=J_Z(\tau)/J_c$ is the normalized bias current density and 
\vskip -8pt
\begin{equation}
F_h(\tau)\equiv h\Delta \frac{\cos\bar{\theta}\cosh\bar{\nu} \sin\tau-\sin\bar{\theta}\sinh\bar{\nu}\cos\tau }{\mathcal{Q}^2(\tau)}
\label{Fh}
\end{equation}
\noindent is an additional forcing term proportional to the applied magnetic field; $h\equiv H/J_c c$ is the normalized field strength for treating long CAJTJs and $\Delta$ is a geometrical factor which sometimes has been referred to as the coupling between the external field and the flux density of the junction \cite{gronbech}. As usual, the $\alpha$ and $\beta$ terms in Eq.(\ref{psge}) account for, respectively, the quasi-particle shunt loss and the surface losses in the superconducting electrodes. Eq.(\ref{psge}) is supplemented by periodic boundary conditions \cite{PRB96}:
\vskip -8pt
\begin{subequations}
\begin{eqnarray} \label{peri1}
\phi(\tau+2\pi,\hat{t})=\phi(\tau,\hat{t})+ 2\pi n,\\
\phi_\tau(\tau+2\pi,\hat{t})=\phi_\tau(\tau,\hat{t}),
\label{peri2}
\end{eqnarray}
\end{subequations}
\vskip -4pt
\noindent where $n$ is an integer number, called the winding number, corresponding to the algebraic sum of \Jos vortices (or fluxons) trapped in the \jun due to flux quantization in one of the superconducting electrodes. Eqs.(\ref{psge}) can be classified as a perturbed and modified sine-Gordon equation in which the perturbations are given by the system dissipation and driving fields, while the modification is represented by an effective local $\pi$-periodic \Jos penetration length, $\Lambda_J(\tau)\equiv \lambda_J/Q(\tau)= c \lambda_J \Delta \nu /\Delta\!W(\tau)$, inversely proportional to the annulus width. It is worth to point out that this $\Lambda_J$ variation stems from the variable junction width and cannot be modeled in terms of a spatially varying $\lambda_J$ in uniform-width JTL treated in Refs.(\cite{sakai},\cite{petras}); nevertheless, in the time independent case, it happens to be equivalent to a change in the $J_c$ of a uniform-width JTL \cite{semerdzhieva}.

\subsection{Alternative derivation of the equation of motion} 

An alternative derivation of the equation of motion for the \Jos phase of a CAJTJ arises from the theory developed by Goldobin \textit{et al} \cite{goldobin01} for one-dimensional curved variable-width JTLs. According to this theory, although adopting our notations, $\phi(\hat{s},\hat{t})$ satisfies the following non-linear PDE:
\vskip -8pt
\begin{equation}
\phi_{\hat{s}\hat{s}} - \phi_{\hat{t}\hat{t}}-\sin \phi =\gamma+ \alpha \phi_{\hat{t}} + \frac{1}{J_c \lambda_J} \frac{dH_\nu}{d\hat{s}}+ \frac{\Delta w_{\hat{s}}}{\Delta w}\left[\frac{H_\nu}{J_c \lambda_J}-\phi_{\hat{s}}\right],
\label{goldobin}
\end{equation}
\vskip -4pt
\noindent where $\hat{s}=s/\lambda_J$. $H_\nu$ is the component of the applied magnetic field normal to the junction perimeter. $\Delta w_s$ is the directional derivative of the local \jun width (for the sake of simplicity, the surface losses were neglected in Ref.\cite{goldobin01}). Eq.(\ref{goldobin}) can be rearranged as:
\vskip -8pt
\begin{equation}
\phi_{\hat{s}\hat{s}} +\frac{\Delta w_{\hat{s}}}{\Delta w} \phi_{\hat{s}} - \phi_{\hat{t}\hat{t}}-\sin \phi =\gamma+ \alpha \phi_{\hat{t}} + \frac{1}{J_c \lambda_J} \left[ \frac{dH_\nu}{d\hat{s}}+ \frac{\Delta w_{\hat{s}}}{\Delta w} H_\nu\right].
\label{goldobin2}
\end{equation}
\vskip -4pt
\noindent In force of Eq.(\ref{nu}), we have:
\vskip -8pt
\begin{equation}
H_\nu(\tau)={\bf {H}}\cdot{\bf \hat{N}}=\frac{H}{\mathcal{Q}(\tau)} \left(\sin\bar{\theta}\sinh\bar{\nu}\sin\tau + \cos\bar{\theta}\cosh\bar{\nu}\cos\tau \right).
\label{NormField} 
\end{equation}
\vskip -4pt
\noindent We stress that, for CAJTJs in a uniform field, the field radial component is very smooth, at variance with the other proposed geometries in which the $\delta$-like behavior of $H_\nu$ makes the modeling only qualitative \cite{goldobin01,kemp02}. Exploiting the fact that, in elliptic coordinates, $\Delta w_{\hat{s}}/\Delta w =-\lambda_J \sin2\tau/2c \mathcal{Q}^3(\tau)$ and considering that
\vskip -8pt
$$\frac{d H_\nu}{d \hat{s}}=\frac{\lambda_J}{c \mathcal{Q}}\frac{d H_\nu} {d\tau}= \frac{H\lambda_J\sinh2\bar{\nu}}{2c}\,\frac{\sin\bar{\theta}\cosh\bar{\nu}\cos\tau -\cos\bar{\theta}\sinh\bar{\nu}\sin\tau}{\mathcal{Q}^4(\nu,\tau)},$$
\vskip -4pt
\noindent after some algebra, one finds that:
\vskip -8pt
\begin{equation}
\label{effeacca} 
 \frac{dH_\nu}{d\hat{s}} + \frac{\Delta w_{\hat{s}}}{\Delta w} H_\nu=J_c \lambda_JF_h(\tau).
\end{equation}
\vskip -4pt
\noindent Furthermore, along the perimeter of a CAJTJ, it is:
\vskip -8pt
\begin{equation}
\label{laplacian} 
\frac{d^2}{d\hat{s}^2}+ \frac{\Delta w_{\hat{s}}}{\Delta w} \frac{d}{d\hat{s}} = \left(\frac{\lambda_J}{c \mathcal{Q}}\right)^2\frac{d^2}{d\tau^2};
\end{equation}
\vskip -4pt
\noindent in passing,we observe that in elliptic coordinates the term proportional to the first spatial derivative has disappeared because $\Delta \nu$ is by definition constant for the confocal elliptic annulus. Inserting the Eqs.(\ref{effeacca}) and (\ref{laplacian}) in Eq.(\ref{goldobin2}), we recover the non-linear PDE in Eq.(\ref{psge}), as well as the magnetic forcing term in Eq.(\ref{Fh}).

\subsection{The Lagrangian and Hamiltonian densities}	

In the absence of dissipation ($\alpha=\beta=0$) and assuming a uniform current distribution, that is, $\gamma(\tau)=\gamma_0$, Eq.(\ref{psge}) can be rewritten as:	

%\begin{equation}
%\left[\frac{\lambda_J}{c\,\mathcal{Q} (\tau)}\right]^2 \phi_{\tau\tau} - \phi_{\hat{t}\hat{t}}=\sin \phi - \left[\frac{\lambda_J}{c\,\mathcal{Q} (\tau)}\right]^2 \kappa H c \Delta \left(\sin\bar{\theta}\sinh\bar{\nu}\cos\tau-\cos\bar{\theta}\cosh\bar{\nu} \sin\tau\right)=\sin \phi - \left[\frac{\lambda_J}{c\,\mathcal{Q} (\tau)}\right]^2 \frac{du_h}{d\tau},
%\label{psge5b}
%\end{equation}

\vskip -8pt
\begin{equation}
 \left(\frac{\lambda_J}{c}\right)^2 \phi_{\tau\tau} - \mathcal{Q}^2(\tau)(\phi_{\hat{t}\hat{t}}+ \sin \phi)=\,-\, \frac{d(u_h+u_\gamma)}{d\tau},
\label{psge1}
\end{equation}
\vskip -4pt
\noindent were we have introduced the functions:
\vskip -8pt
\begin{equation}
u_h(\tau)\equiv\frac{H_\nu(\tau)	\mathcal{Q}(\tau)\Delta}{J_c c}=h \Delta \left(\sin\bar{\theta}\sinh\bar{\nu}\sin\tau +\cos\bar{\theta}\cosh\bar{\nu} \cos\tau \right),
\label{uh}
\end{equation}
\vskip -4pt
%\noindent [[$=h \Delta\, q(\bar{\nu},\bar{\theta}) \cos(\xi-\tau)$. Here, we introduced the angle $\xi(\bar{\nu},\bar{\theta})$ such that $\sin\xi=\sin\bar{\theta} \sinh\bar{\nu} /q(\bar{\nu}, \bar{\theta})$ and $\cos\xi=\cos\bar{\theta} \cosh\bar{\nu} /q(\bar{\nu}, \bar{\theta})$. In passing, we observe that $\tan \xi=\tan\bar{\theta} \tanh\bar{\nu}$. In Eq.(\ref{uh}) the potential zero was set in $\tau=\xi\pm \pi/2$.]]

%[[$h_\nu(\hat{s})d\hat{s}=u_h(\tau)d\tau/\Delta$]]
\noindent  and
\vskip -8pt
\begin{equation}
u_\gamma(\tau)\equiv \frac{\gamma_0}{2}\left( \tau \cosh2\bar{\nu} + \frac{1}{2}\sin2\tau\right),
\label{ugamma}
\end{equation}
\vskip -4pt
\noindent proportional, respectively, to the magnetic field, $h$, and the bias current, $\gamma_0$; their physical meaning will be given later on. For the time being, we just recognize that $du_h /d\tau=F(\tau) \mathcal{Q}^2(\tau)$ and $du_\gamma /d\tau=\gamma_0 \mathcal{Q}^2(\tau)$. Eq.(\ref{psge1}), although not integrable (unless when $\mathcal{Q}(\tau)=\textrm{const}$ and $u_h=u_\gamma=0$, in which case it is the well known sine-Gordon equation), admits (numerically computed) solitonic solutions \cite{JLTP16b}; nevertheless, in analogy with the pure sine-Gordon equation, it may be derived by substituting the Lagrangian density: 
\vskip -8pt
\begin{equation}
\hat{\mathcal{L}}(\tau,\hat{t},\phi,\phi_\tau,\phi_{\hat{t}})=\frac{1}{2} \frac{\lambda_J}{c} \left( \phi_\tau + u_h + u_\gamma \right)^2 + \mathcal{Q}^2(\tau) \frac{c}{\lambda_J} \left(-\frac{1}{2} \phi^2_{\hat{t}}+1-\cos\phi \right)
\label{lagrangian}
\end{equation}
\vskip -8pt
\noindent into the Euler-Lagrange equation:
\vskip -8pt
$$\frac{d}{d \tau} \left( \frac{\partial \hat{\mathcal{L}}}{\partial \phi_\tau}\right)+ \frac{d}{d{\hat{t}}} \left( \frac{\partial \hat{\mathcal{L}}}{\partial \phi_\tau}\right)  - \frac{\partial \hat{\mathcal{L}}}{\partial \phi}=0.$$
\vskip -4pt
\noindent Then, the Hamiltonian density is \cite{note1}:
\vskip -8pt
\begin{equation}
\hat{\mathcal{H}}(\tau,\hat{t},\phi,\phi_\tau,\phi_{\hat{t}})\equiv \hat{\mathcal{L}}+\phi_{\hat{t}}\frac{\partial \hat{\mathcal{L}}}{\partial \phi_{\hat{t}}}=\frac{1}{2} \frac{\lambda_J}{c} \left( \phi_\tau + u_h +u_\gamma \right)^2 + \mathcal{Q}^2(\tau)\frac{c}{\lambda_J} \left(\frac{1}{2} \phi^2_{\hat{t}}+1-\cos\phi \right),
\label{hamiltonian}
\end{equation}
\vskip -4pt
\noindent and, the system total energy, i.e., the Hamiltonian is: 
\vskip -8pt
\begin{equation}
\hat{E}=\int^{\pi}_{-\pi}\hat{\mathcal{H}}\, d\tau.
\nonumber
\label{energy}
\end{equation}
\vskip -4pt
\noindent $\hat{E}$ is an integral of motion in the absence of bias current; in fact, using Eq.(\ref{psge1}) and owing to the periodicity of $\phi$ and $u_h$, one obtains: $d\hat{E}/d{\hat{t}}= \left[ \phi_{\hat{t}}\left( \phi_\tau + u_h + u_\gamma \right) \right]^{\pi}_{-\pi}=0$.

%$$\frac{d\hat{H}}{d{\hat{t}}}= \left(\frac{\lambda_J}{c}\right) \left[ \phi_{\hat{t}}\left( \phi_\tau + u_h + u_\gamma \right) \right]^{\pi}_{-\pi}=0.$$
 
%\noindent The right-hand side of Eq.(\ref{}) is usually considered as a perturbation \cite{scott}; it does not drastically change the vortex shape, but defines its dynamics, e.g., its equilibrium velocity. Such an approximation essentially treats the vortex as a rigid object, and its dynamics can be reduced to that of a relativistic underdamped point-like particle \cite{carapella,goldobin12}.

\subsection{The approximate solitonic solution}

%Any static solution ($\tilde{\phi}_{\hat{t}}=0$) of  Eq.(\ref{psge1}) must satisfy the equation:
%\vskip -8pt
%\begin{equation}
%\left(\frac{\lambda_J}{c}\right)^2 \phi_{\tau\tau} - \mathcal{Q}^2(\tau) \sin \phi = -\frac{d(u_h+u_\gamma)}{d\tau}.
%\label{static}
%\end{equation}

\noindent In the absence of the right-hand side, the simplest solitonic solution of Eq.(\ref{psge1}) on an infinite line, in a first approximation, is a single Josephson vortex (sine-Gordon kink) centered at $s_0(\hat{t})$ and moving with instantaneous (tangential) velocity $\dot{s}_0=ds_0/d\hat{t}$, namely, $\tilde{\phi}(\tau,\hat{t})= 4 \arctan \exp \left\{\wp[s(\tau)-s_0(\hat{t})]/\lambda_J \right\}$, where $\wp=\pm1$ is the fluxon polarity \cite{scott}. Indeed, the phase profile:
%[[This is what is commonly done!]]
\vskip -8pt
\begin{equation}
\tilde{\phi}(\tau,\hat{t})= 4 \arctan \exp \left\{ \wp \left[ \frac{c \cosh\bar{\nu}}{\lambda_J} \texttt{E}(\tau,e^2)- \frac{s_0(\hat{t})}{\lambda_J} \right]\right\},
\label{tilde}
\end{equation}
\vskip -4pt
%[[see 4ArcTanExpNew.nb]]
\noindent satisfies the left-hand side of Eq.(\ref{psge1}), provided that \cite{note2} $c/{\lambda_J} >>1/\sinh^3\bar{\nu}$ and both the normalized fluxon speed, $\hat{u}\equiv\dot{s}_0/{\lambda_J}$, and acceleration, $\hat{a}\equiv \ddot{s}_0/{\lambda_J}$, are much less than unity (in moduli). Since $(1/2)(\lambda_J/c)^2 \tilde{\phi}_\tau^2 =\mathcal{Q}^2 (1-\cos\tilde{\phi})$, then the inductive energy density equals the Josephson energy density; more specifically, with $\mathcal{Q}(\tau)d\tau=(1/c)ds$, it is:
\vskip -8pt
$$\frac{1}{2} \frac{\lambda_J}{c} \int^{\infty}_{-\infty} \tilde{\phi}_\tau^2\, d\tau=\frac{c}{\lambda_J} \int^{\infty}_{-\infty} \mathcal{Q}^2(\tau)(1-\cos\tilde{\phi})\, d\tau=$$
\vskip -4pt
$$=\frac{2}{\lambda_J} \int^{\infty}_{-\infty} \mathcal{Q}(s) \sech^2\!\frac{\wp(s-s_0)}{\lambda_J} ds \approx 4 \mathcal{Q}(\tau_0).$$
\vskip -4pt
\noindent Here we used the convolution integral identity $\int^{\infty}_{-\infty} f(x)\sech^2[\pm(x-x_0)/p] dx\approx 2p f(x_0)$, if $p <<1$, and the fact that the width of a long CAJTJ does not change much over a distance compared to the fluxon size. Therefore, according to the expression of the energy density in Eq.(\ref{hamiltonian}) (disregarding for the time being the magnetic field and the bias current), the potential energy periodically depends on its position through the scale factor Q which also describes the smooth position-dependent width of a CAJTJ (see Eq.(\ref{width})):
\vskip -8pt
\begin{equation}
\hat{U}_0(\tau_0) \approx 8 \mathcal{Q}(\tau_0)= 8 \frac{\Delta w(\tau_0)}{c \, \Delta \nu}.
\label{U0}
\end{equation}

\noindent As anticipated in the Introduction, the existence of a fluxon repelling (attracting) barrier is induced by a widening (narrowing) JTL, as first first reported by Nappi and Pagano \cite{nappipagano} and later on by other authors \cite{benabdallah96,goldobin01,kemp10}. For CAJTJs, Eq.(\ref{U0}) expresses a $\pi$-periodic potential uniquely determined by the annulus ellipticity, $e^2\equiv \sech^2\bar{\nu}$. The potential wells are located at $\tau_0=\pm \pi/2$, where the annulus width is smallest (see the solid line in Figure~\ref{pots}). Considering that $\sinh\bar{\nu}\leq \mathcal{Q}(\tau) \leq \cosh\bar{\nu}$, the potential wells are separated by an energy barrier proportional to the exponential of $\bar{\nu}$. We stress that $\hat{U}_0$ is an intrinsic potential, i.e., it occurs in the absence of an applied magnetic field. Indeed, it differs from the sinusoidal potential induced by a small uniform field applied to a circular annular JTL under several aspects: i) $\hat{U}_0$ has an halved periodicity, i.e., there are two minima and two maxima for every round trip; ii) $\hat{U}_0$ is proportional to $\phi_\tau^2$ and so is independent on the fluxon polarity, while a magnetic potential complies with the fluxon polarity; iii) by squashing the annulus the relative inter-well barrier height can be made arbitrarily large, albeit limited by the resolution of the lithographic processes and the accuracy of the mask alignment during the fabrication process.

\begin{figure}[tb]
\centering
%\subfigure[ ]{\includegraphics[width=7cm]{.png}}
%\subfigure[ ]{\includegraphics[width=7cm]{.png}}
\includegraphics[width=8cm]{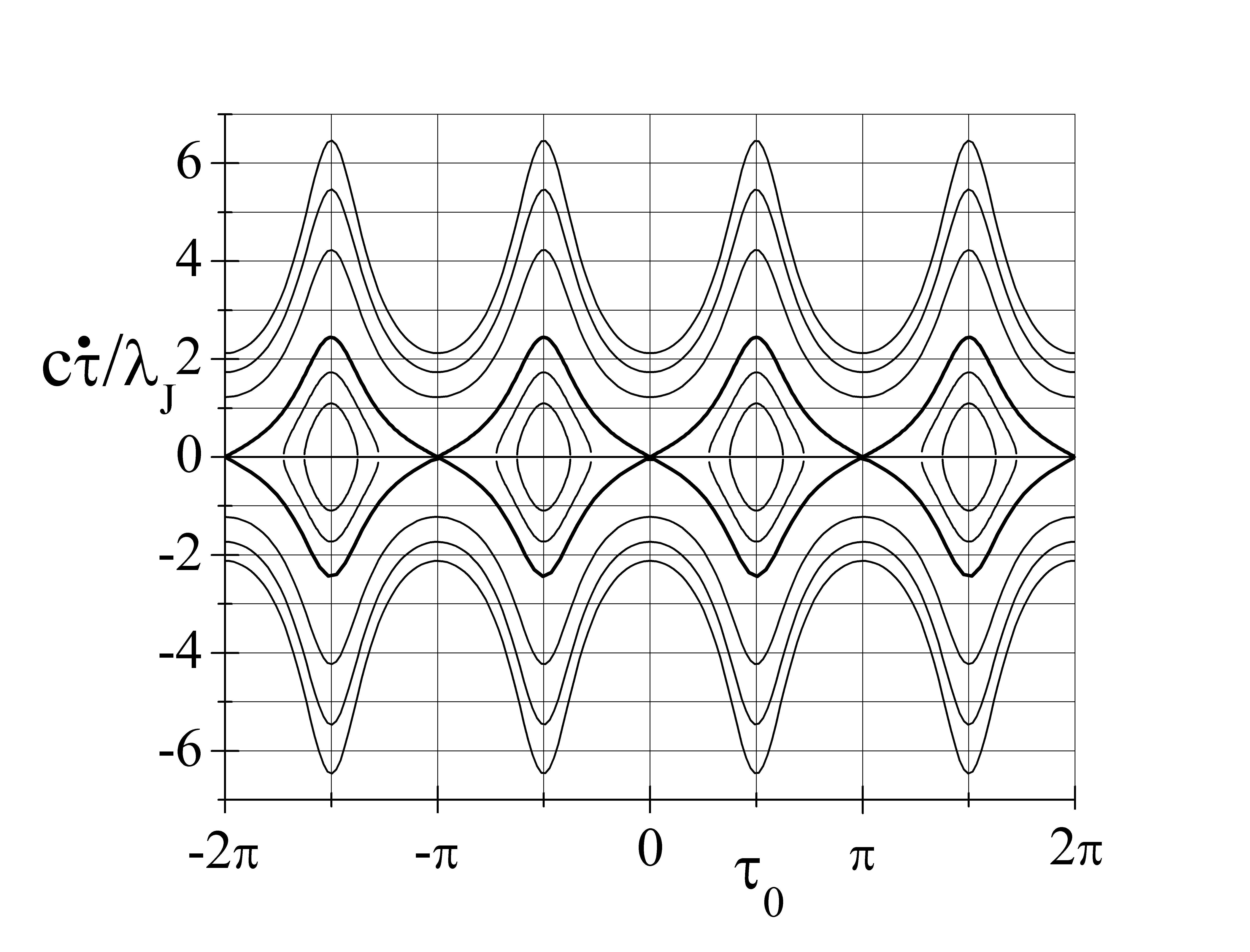}
\caption{Calculated phase portrait for a fluxon in a CAJTJ when $\bar{\nu} = 0.55$. Each curve represents a different energy. The thick line is the separatrix curve for $\epsilon=\coth \bar{\nu}$ separating the two dynamic states. The fluxon rest energy correspond to $\epsilon=1$.}
\label{phasespace}
\end{figure}
\noindent The integration of the $\phi^2_{\hat{t}}$ term in Eq.(\ref{hamiltonian}) yields a kinetic energy, $\hat{K}\equiv \hat{U}_0 \hat{u}^2/2$, suggesting that $\hat{U}_0$ can also be interpreted as the fluxon position-dependent mass. The fluxon normalized energy is then given by:
\vskip -8pt
\begin{equation}
\hat{E} =\hat{K}+ \hat{U}_0= \hat{U}_0 (1+\frac{1}{2}\hat{u}^2).
\label{hatE}
\end{equation}

\noindent This equation replaces the expression for the fluxon normalized energy given in the Introduction for a constant-width JTL provided that $c \Delta\nu$ is put in place of $W$ in the characteristic energy $\mathcal{E}$. We might extrapolate Eq.(\ref{hatE}) to relativist velocities as $\hat{E} = \hat{U}_0/\sqrt {1-\hat{u}^2}$; we believe that this expression holds for any potential induced by the width variation. The minimum energy corresponds to a static fluxon ($u=0$) pinned in one of the potential wells at $\tau_0=\pi/2+k\pi$, i.e., $\hat{U}_0(\pm \pi)=8\sinh\bar{\nu}$ also represent the fluxon rest energy $\hat{E}_{rest}$. We can express the energy as $\hat{E}=\epsilon \hat{E}_{rest}$ with $\epsilon\geq 1$, so that:
\vskip -8pt
\begin{equation}
\hat{u}=\pm \sqrt{2\left( \frac{\epsilon \sinh \bar{\nu}}{Q}-1\right)}.
\label{hatu}
\end{equation} 
\vskip -6pt
%[[see Calculations/PhaseSpace.pdf]]
\noindent Therefore, $\epsilon$ can be used to parametrize the fluxon trajectories in the phase plane, as shown in Figure~\ref{phasespace} for $\bar{\nu} = 0.55$ ($e^2=3/4$), where each curve represents a different energy. Low energies correspond to bound states in which the fluxon oscillates around one of the equilibrium points. For higher energies the fluxon propagates with a modulated speed. The two sets of dynamical states are set apart by the separatrix curve corresponding to $\epsilon = \coth \bar{\nu}$, that is, to $\hat{E}=8\cosh\bar{\nu}$. Very energetic, i.e., relativistic vortices, not represented in the figure, are barely affected by the periodic potential and can move ballistically in an almost force-free environment, $\hat{u}^2=1-\hat{U}_0^2/\hat{E}^2$. 

\subsection{The quantum regime}

\noindent According to quantum mechanics, a massive particle subjected to potential confinement has its energy quantized and a discrete energy spectrum would be expected in the classical region of positive kinetic energy. Recently, macroscopically distinct quantum states of a vortex trapped in a magnetic field controlled double-well potential inside a narrow long heart-shaped junction \cite{kemp02} and in a linear JTL with local field injectors \cite{fistulPRB03} have been used for designing qubits. In this paragraph we investigate the conditions under which the quantum-mechanical regime arises in CAJTJs. To calculated the allowed energy levels $E_n$ of a non-relativistic particle moving in a field of potential energy given in Eq.(\ref{U0}), we first focus on the bound states with the lowest energies. Any potential $U(q)$ is approximately harmonic, i.e., parabolic, in the neighborhood of a local minimum $q_0$, $U(q)\approx (1/2)U''(q_0)(q-q_0)^2$, since $U(q_0)$ is a constant that does not change the force and, by definition, $U'(q_0)=0$; therefore, the ground state energy and the lowest excited energies are well given by the harmonic oscillator eigenenergies, $E_n=(n+1/2)\hbar \omega_0$, where $\omega_0=2\pi/P_0$ and $P_0$ is the classical period of small-amplitude, i.e., small-energy, oscillations. Since for an harmonic oscillator the oscillation period does not depend on the oscillation amplitude, the anharmonicity degree of a potential well is measured by the energy-dependence of the oscillation period:
\vskip -8pt
$$P(E)=\oint dt=2\int_{q_1(E)}^{q_2(E)} \frac{dq}{u(E)},$$
\vskip -6pt
\noindent where $u$ is the particle velocity, $q_1$ and $q_2$ are the turning points where the kinetic energy vanishes, namely, $U(q_1)=U(q_2)=E$ and the factor $2$ takes into account the back and forth motion. Making use of the expression for the velocity in Eq.(\ref{hatu}), for our potential the anharmonic oscillation period is:
\vskip -6pt
\begin{equation}
P(\epsilon,\bar{\nu})=\sqrt{2} \frac{c}{\bar{c}} \int_{\tau_-}^{\tau_+} \frac{Q^{3/2}}{\sqrt{\epsilon \sinh \bar{\nu}-Q}}\,d\tau ,
\label{period}
\end{equation} 
\vskip -8pt
 \begin{figure}[t]
\centering
%\subfigure[ ]{\includegraphics[width=7cm]{.png}}
%\subfigure[ ]{\includegraphics[width=7cm]{.png}}
\includegraphics[width=8cm]{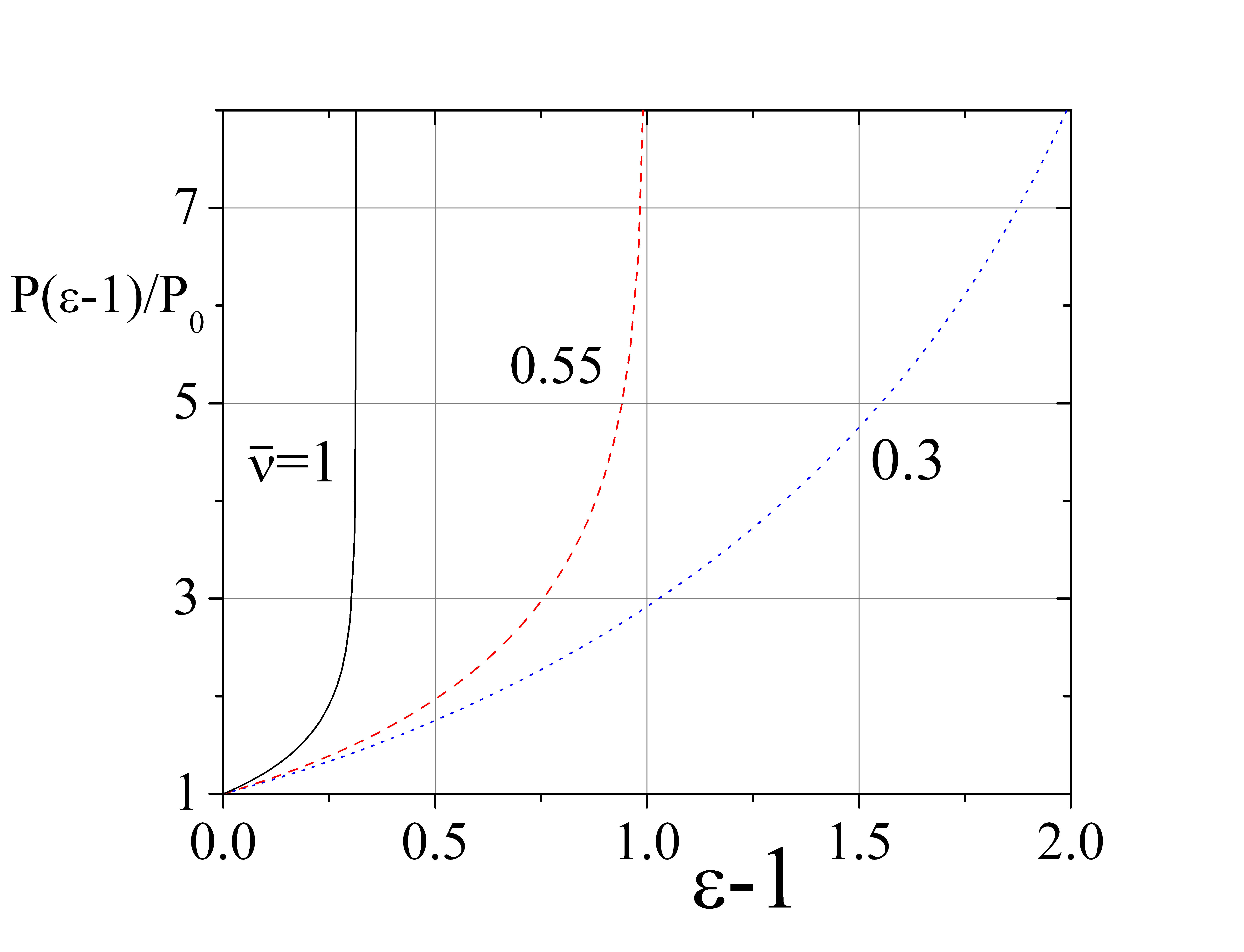}
\caption{(Color online) Energy dependence of the normalized small amplitude oscillation period in Eq.(\ref{period}) for three values of $\bar{\nu}$.}
\label{integral}
\end{figure}
%[[see Calculations/Oscillation period.pdf avd NoteBooks/Oscillation period.nb]]
\noindent where $\tau_\pm(\epsilon,\bar{\nu})=\pi/2 \pm \arcsin (\sqrt{\epsilon^2-1}\sinh \bar{\nu})$. The definite integral in Eq.(\ref{period}) has been evaluated numerically and $P(\epsilon)$ is shown in Figure~\ref{integral} for three values of $\bar{\nu}$. We see that $P(\epsilon-1)$ increases linearly for small energies and eventually diverges when the energy equals the potential maximum; in other words, as expected for a finite-wall potential, it takes an infinite time for a particle to oscillate along the separatrix curve of the phase space. It was also found that, in the small amplitude limit ($\epsilon \to 1$), it is $P_0(\bar{\nu})\equiv P(1,\bar{\nu})=2\pi (c/\bar{c})\sinh^2 \bar{\nu}$, so that the angular frequency is $\omega_0(\bar{\nu})\equiv 2\pi/P_0(\bar{\nu}) =\bar{c}/c \sinh^2 \bar{\nu}$. The same result could be obtained using the definition $\omega_0 \equiv \sqrt{U''(q_0)/m(q_0)}$; furthermore, being $\bar{c} =\omega_{p}\lambda_J$, we have that the harmonic frequency is proportional to the plasma frequency, $\omega_0(\bar{\nu})= \omega_{p}\lambda_J /c \sinh^2 \bar{\nu}$. We note that, being independent on the junction width, $\omega_0$ does not depend on the fluxon mass, at variance with the case of the harmonic oscillator of mass $m'$ and spring constant $k'$ for which $\omega_{harm}^2=k'/m'$.
%[[see and Calculations/omega0.pdf]]In classical mechanics, anharmonicity is the deviation of a system from being a harmonic oscillator.
The quantum regime settles in when the separation, $\hbar \omega_0$, between these energy levels becomes much larger than the thermal energy, $k_B T$. Assuming a Swihart velocity about 30 times smaller than the speed of the light in the vacuum, a distance $2c$ between the CAJTJ foci of about $200\, \mu m$ and a moderate eccentricity $\bar{\nu}=0.55$, we have that for an (unbiased) CAJTJ $\omega_0\approx 300 GHz$ falls in the mm-wave domain which, in the limit of small damping, results in a crossover temperature \cite{devoret85, cirillo16} $\hbar \omega_0/7.2 k_B\approx 300\,\mu K$; this estimate results to be consistent with the transition from thermal to quantum regime observed at $200\, mK$ by Wallraff \textit{et al.} \cite{wallraff03} in a biased ring-shaped JTL under an external magnetic field (the discrepancy may be accounted for by the dependence of the angular frequency on the normalized bias current). In the lowest region of the energy spectrum the probability that the fluxon tunnels to a neighboring degenerate well is extremely small since the barrier potential is very large and very wide. In order to find the higher eigenenergies of the bounded states, we resort to the Bohr and Sommerfeld's quantization rule according to which \cite{landau}, even for large quantum numbers $n$, the distance between two neighboring levels is $\Delta E=h/P(\epsilon,\bar{\nu})=\hbar \omega(\epsilon,\bar{\nu})$. From Figure~\ref{integral} we see that $P$ increases, and eventually diverges, with the energy, so that the level separation decreases until the energy spectrum becomes continuous when the fluxon eventually turns to be unbound (depinned) for $\epsilon>\coth\bar{\nu}$. Although the energy levels are not equidistant, the angular frequencies $\omega$ can be regarded as approximately the same for several adjacent levels. 
%The reduced or normalized Plank's constant $\hbar \omega_p/E_0$ was considered in kato&imada, wallraff00, kemp10 amd fujii08
\noindent It is worth to stress that the eigenenergies, $E_n=\epsilon_n E_{rest}$, are measured from the bottom of the potential that is the fluxon rest energy, $E_{rest}$. For the ground state ($n=0$), it must be $(\epsilon_0-1)E_{rest} =\hbar \omega_0 /2$. Sometimes in the literature \cite{kato96, fujii08} the prefactor $(\epsilon_n-1)$ has been omitted and the condition $E_{rest}/ \omega_p =O(\hbar)$ was invoked to determine the maximum JTL width (typically tens to hundreds of nanometers) that would allow to enter the quantum regime. We believe that this very restrictive condition is not needed, with $\epsilon_0$ (and the first few $\epsilon_n$) sufficiently close to the unity. To be more precise on this point we consider that the semi-classical quantization requires the linear (non-relativistic) momentum, $p=mu$, of a particle confined in a single minimum potential, to be small enough so that the path integral, $\mathcal{A}=\oint p\,dq$, along a classical oscillating trajectory is a half-integer number of quanta of action. While for the harmonic oscillator it is $\mathcal{A}_{harm} = 2\pi E/\omega_{harm}$, for our potential it is: 
\vskip -8pt
\begin{equation}
\mathcal{A}(\epsilon,\bar{\nu})=8\sqrt{2} \frac{\mathcal{E}}{\omega_p} \int_{\tau_-}^{\tau_+} Q^{3/2}\sqrt{\epsilon \sinh \bar{\nu}-Q}\,d\tau ,
\label{action}
\end{equation} 
%[[see Calculations/BohrSommerfeldQuantisation.pdf]]
\noindent where $\mathcal{E}=\Phi_0 J_c \lambda_J c \Delta\nu/2\pi$. Numerical analysis shows that the integral in Eq.(\ref{action}) starts from zero for $\epsilon=1$ and increases linearly with $\epsilon-1$, but never exceeds the value of $\sqrt{2} \sinh \bar{\nu}$. For ($\epsilon \to 1$), we have 
\vskip -8pt
$$\mathcal{A}(\epsilon \to 1,\bar{\nu})\approx 16 \pi \frac{\mathcal{E}} {\omega_p} (\epsilon -1) \sinh^3\bar{\nu}= 2\pi\frac{E-E_{rest}}{\omega_0}.$$ 
\vskip -4pt
\noindent and, more generally,
\vskip -8pt
$$\mathcal{A}(\epsilon,\bar{\nu})\approx 2\pi \frac{(\epsilon-1) E_{rest}}{\omega(\epsilon,\bar{\nu})}.$$ 
\vskip -4pt
%\noindent Therefore, imposing, by way of example, $\mathcal{A}=10h$, provides a maximum value for the junction width, $\overline{\Delta W}\lesssim 0.1 \csch\bar{\nu}\, \mu m$, that, for $\bar{\nu}=0.55$, yields $\overline{\Delta W}\lesssim 0.2\, \mu m$. 

\noindent As the annulus width is reduced, the fluxon rest mass decreases and, in turn, the relative separation between the discrete energy levels increases so that the quantum regime persists to higher temperature. It is worth to mention that, as the JTL width becomes comparable to the electrodes London penetration depth, a significant fraction of the fluxon energy lies outside the junction which makes the specific characteristic energy, $\mathcal{E}/W$ (or, in our case $\mathcal{E}/c\Delta\nu$), to decrease with decreasing junction width \cite{ustinov06}. All the above considerations suggest that quantum effects become more pronounced as $\bar{\nu}$ decreases which corresponds to an increase of the CAJTJ eccentricity. 

\section{The double-well potential}

\subsection{The perturbed potential}

\noindent The right-hand side of Eq.(\ref{psge1}) is usually considered as a perturbation \cite{scott} that does not drastically change the vortex profile. Therefore, inserting the vortex profile Eq.(\ref{tilde}) in the Hamiltonian density Eq.(\ref{hamiltonian}), one can derive the explicit expression of the potential energy, $\hat{U}(\tau_0)$, as a function of the fluxon coordinate $\tau_0$. Recalling that $\int^{\infty}_{-\infty} f(x) \sech[\pm(x-x_0)/p] dx\approx\pi p f(x_0)$, if $p <<1$, when we take into account the magnetic field and the bias current, the integration of $\hat{\mathcal{H}}$ in Eq.(\ref{hamiltonian}) with $\phi= \tilde{\phi}(\tau)$ yields two extra tunable terms which build the fluxon external potential, namely, the $2\pi$-periodic magnetic potential, $\hat{U}_h(\tau_0)\approx 2\pi \wp (\lambda_J/c) u_h(\tau_0)$, and the current-induced tilting potential, $\hat{U}_\gamma(\tau_0)\approx 2\pi \wp (\lambda_J/c) u_\gamma(\tau_0)$. $\hat{U}_h$ is $\pi$-antiperiodic in $\tau$, i.e., $\hat{U}_h(\tau+\pi) =-\hat{U}_h(\tau)$, then it averages to zero over one period. For a \Jos ring, with $\tau$ replaced by $\theta$ and $\bar{\nu}\to \infty$, we recover the sinusoidal magnetic potential \cite{PRB97}, $\hat{U}_h(\theta)\propto \cos(\bar{\theta}-\theta)$.

\noindent Resuming, the total potential energy experienced by the vortex along the length of a long CAJTJ is:
\vskip -8pt
\begin{equation}
\hat{U}(\tau_0)=\hat{U}_0(\tau_0)+\hat{U}_h(\tau_0)+\hat{U}_\gamma(\tau_0)\approx  8 \mathcal{Q} + 2\pi \wp \frac{\lambda_J}{c}\left(u_h+ u_\gamma \right).
\label{Utot}
%\nonumber
\end{equation}
\vskip -4pt
\begin{figure}[tb]
\centering
%\subfigure[ ]{\includegraphics[height=8cm,width=7cm]{MDP4Pi.png}}
%\subfigure[ ]{\includegraphics[height=8cm,width=7cm]{empty.pdf}}
\includegraphics[width=8cm]{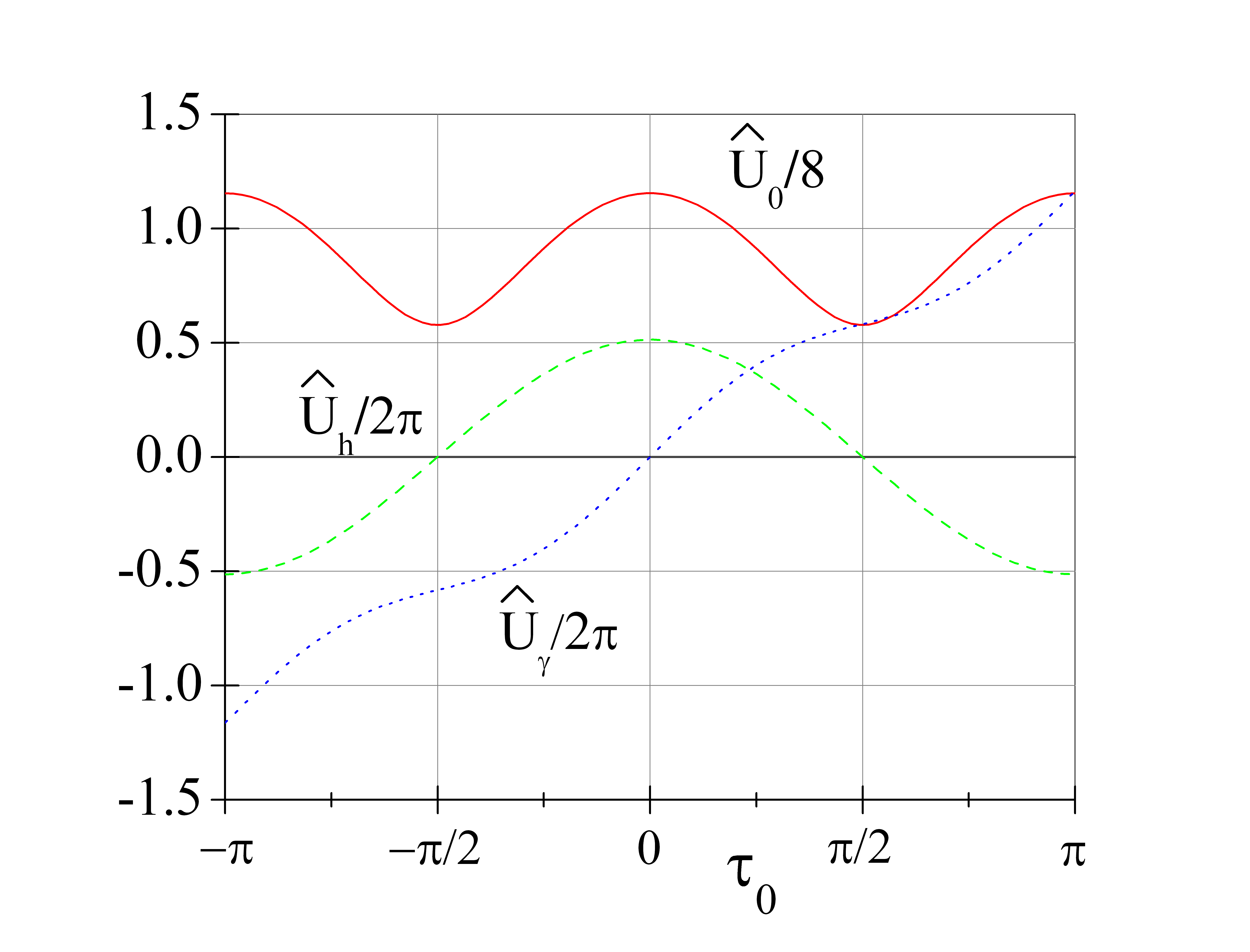}
\caption{(Color online) The three fluxon potentials in Eq.(\ref{Utot}) plotted versus the position of its center of mass, $\tau_0$, for a CAJTJ with $\bar{\nu}=0.55$ (corresponding to an aspect ratio $\rho=1/2$) and $c/\lambda_J=2.25$ (corresponding to a normalized perimeter $\ell=L/\lambda_J=4\pi$): i) the width-dependent potential, $\hat{U}_0$, solid red line, ii) the magnetic potential, $\hat{U}_h$, with $\wp=h=\Delta=1$ and $\bar{\theta}=0$ (dashed green line), and iii) the current potential, $\hat{U}_\gamma$, with $\wp=\gamma_0=1$ (dotted blue line).}
\label{pots}
\end{figure}
%[[see Notebooks/DoubleWellPotential.nb]]
\noindent The three potentials in Eq.(\ref{Utot}) are plotted in Figure~\ref{pots} for $\bar{\nu}=0.55$, $c/\lambda_J=2.25$, $\wp=h=\Delta=\gamma_0=1$ and $\bar{\theta}=0$. Clearly, a large variety of potential can be constructed by tuning, not only the amplitude, but also the orientation $\bar{\theta}$ of the externally applied magnetic field. In addition, the potential profile can be tilted either to left or to right depending on the polarity of the bias current, $\gamma_0$. The inclination is proportional to the Lorentz force acting on the vortex which is induced by the bias current applied to the junction.

\subsection{The washboard potential ($H=0$)}

\begin{figure}[tb]
\centering
%\subfigure[ ]{\includegraphics[width=8cm]{.png}}
%\subfigure[ ]{\includegraphics[width=8cm]{.png}}
\includegraphics[width=8cm]{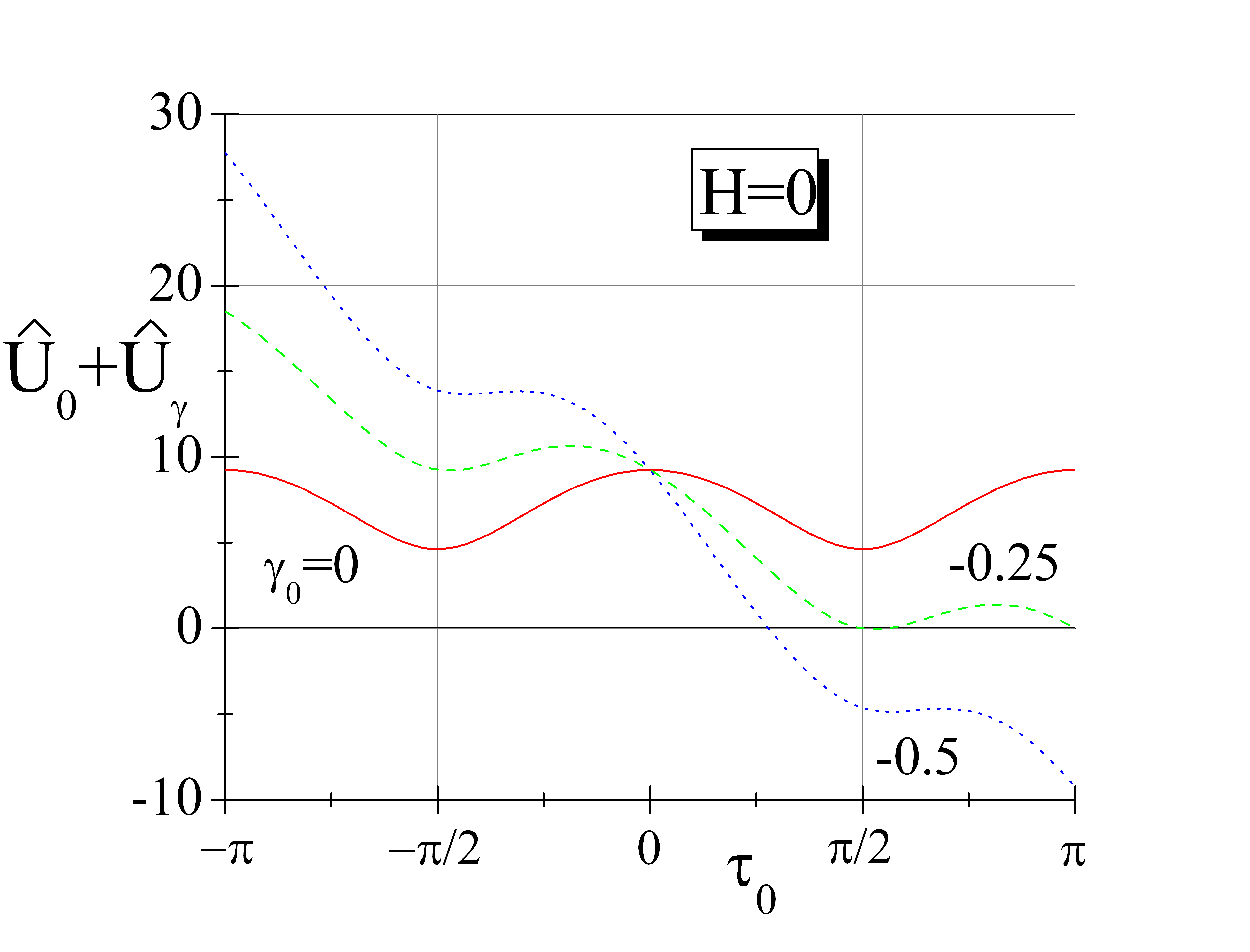}
\caption{(Color online) Tilting of the unperturbed potential $\hat{U}_0$ for $\bar{\nu}=0.55$, $c/\lambda_J=2.25$ and three values of the bias current $\gamma_0$.}
\label{washboard}
\end{figure}

The tilted potential, also called washboard potential, is plotted in Figure~\ref{washboard} for few (negative) values of $\gamma_0$ in the absence of an external magnetic field; as $\gamma_0$ is increased the metastable potential wells get shallower with correspondingly smaller oscillation frequency. This potential is qualitatively similar to the well-studied potential for the phase of a small JTJ biased below its critical current. A particle may escape from a well in the tilted potential by a thermally activated process or by quantum inter-well tunneling. At low temperatures thermal activation is exponentially suppressed, and the escape occurs by macroscopic quantum tunneling \cite{devoret84}. This process can be resonantly activated in presence of a weak microwave perturbation. The existence of quantized levels of the vortex energy within the trapping potential well was demonstrated by measuring the statistics of the vortex escape from a magnetically-induced tilted pinning potential in a $0.5\, \mu m$-wide ring-shaped JTL at temperatures below $100\,mK$ \cite{wallraff00}. Similar experiments can be carried out on a CAJTJs in the absence of an externally applied magnetic field. 

\subsection{The double-well potential with $\gamma=\bar{\theta}=0$}

In the absence of bias current, as soon as we apply a magnetic field, the fluxon unperturbed potential looses the $\pi$-periodicity and becomes $2\pi$-periodic. When the field $h_{\bot}$ is applied perpendicular to the longer annulus diameter, i.e., with $\bar{\theta}=0$ as shown in Figure~\ref{pots}, the potential is still invariant under parity transformation $(\tau_0 \to -\tau_0)$ and develops into a field-controlled symmetric potential with finite walls and two spatially separated minima: 
\vskip -8pt
$$\hat{U}_{DW}(\tau_0)=8 \mathcal{Q} + 2\pi \wp \frac{\lambda_J}{c} u_h + \textrm{const} \approx $$
\vskip -8pt
\begin{equation}
\approx \frac{2\sqrt{2}}{\sqrt{\cosh2\bar{\nu}}}(1+\cos 2\tau_0)+ 2\pi \wp h_{\bot} \frac{\lambda_J}{c} \Delta \cosh\bar{\nu} \cos\tau_0 + \frac{\pi^2 \lambda_J^2 h_{\bot}^2 \cosh^2\bar{\nu} \sqrt{\cosh2\bar{\nu}}}{4 \sqrt{2} c^2}, 
% + \textrm{const}
\label{double}
\end{equation}
\vskip -4pt
%[[see Qmat.nb and ModerateEccen.nb]]
\begin{figure}[tb]
\centering
\subfigure[ ]{\includegraphics[height=5cm,width=7cm]{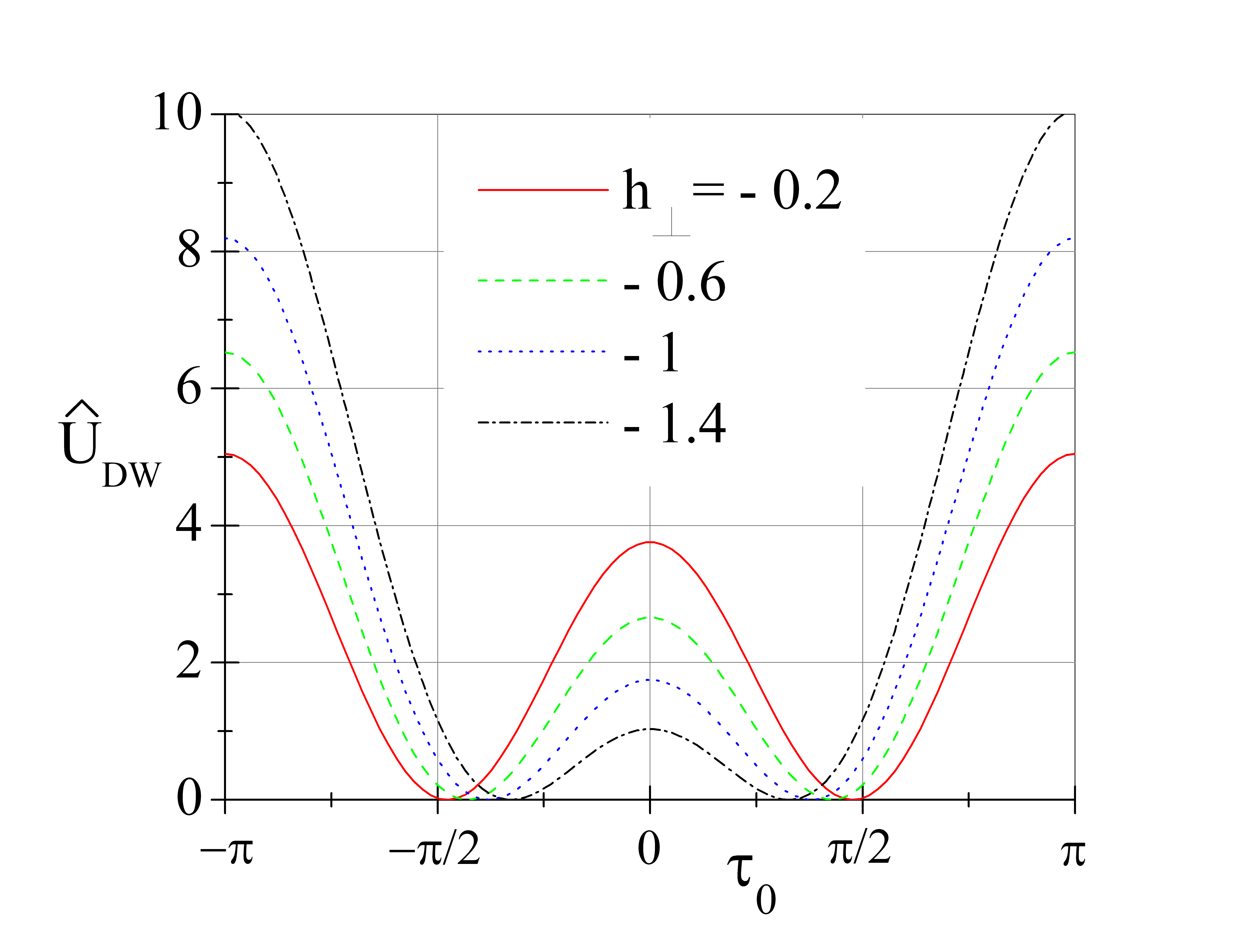}}
\subfigure[ ]{\includegraphics[height=5cm,width=7cm]{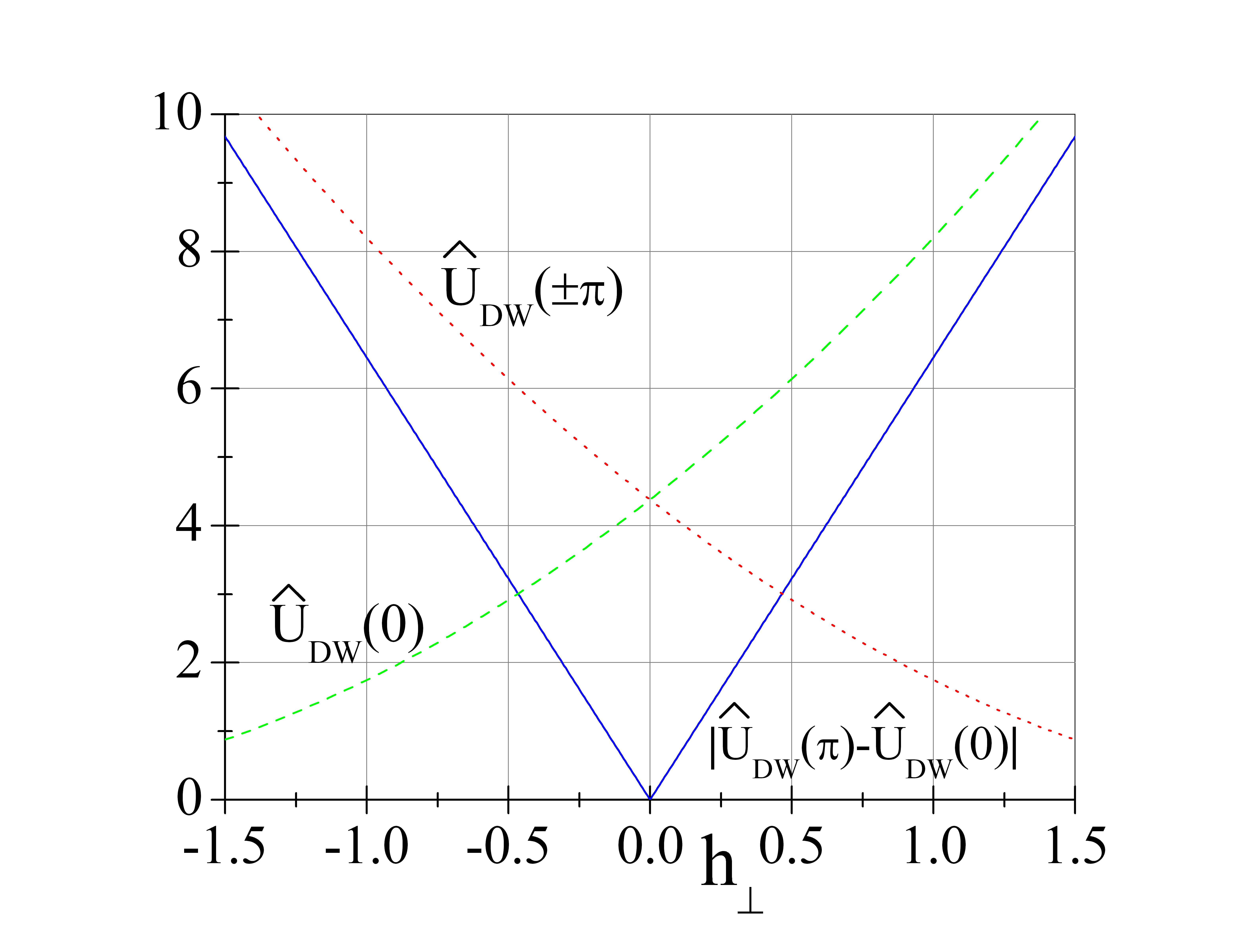}}
\caption{(Color online) (a) The double-well potential as in Eq.(\ref{double}) with $\bar{\nu}=0.55$ and $c/\lambda_J=2.25$, for several negative values of the product $\wp h_{\bot} \Delta$, namely, $-0.2$ red curve, $-0.6$ green, $-1$ blue and $-1.5$ black. (b) The dependence on the perpendicular field, $h_{\bot}$, of i) the inter-well potential $\hat{U}_{DW}(0)$ (green dashed curve), ii) the intra-well barrier height $\hat{U}_{DW}(\pm \pi)$ (red dotted curve), and iii) the absolute value of their difference (blue solid curve). The negative (positive) field values refer to the fluxon (anti-fluxon) potential.}
\label{doublewell}
\end{figure}
%[[see Notebooks/DoubleWellPotential.nb]]
\noindent where a new additive constant has been chosen such that the double-well potential zeroes occur at its minima $\pm \tau_m$ for any values of the parameters, that is, $\hat{U}_{DW}(\tau_m)=0$ with $\tau_m \equiv \text{ArcCos}\left(-{\pi h_{\bot} \lambda_J \cosh\bar{\nu} \sqrt{\cosh2\bar{\nu}}}/{4 \sqrt{2} c}\right)$. Strictly, Eq.(\ref{double}) describes a lattice of double-minima potential. For the sake of simplicity, we have assumed a moderate annulus eccentricity, $e^2\leq3/4$ (corresponding to $\rho\geq 0.5$ and $\bar{\nu}\gtrsim 0.55$), so that $\mathcal{Q}(\tau)$ can be very well approximated by its truncated Fourier expansion, $\mathcal{Q}(\tau)\approx (2/\pi)\cosh\bar{\nu} \,\texttt{E}(e^2) + \cos 2\tau/2\sqrt{2\cosh\,2\bar{\nu}}$, and the unperturbed potential, $\hat{U}_0$, turns into an intrinsic sinusoidal potential whose properties have been well investigated both in the thermal and quantum-mechanical regimes \cite{wallraff00,wallraff03}. The double-well potential in Eq.(\ref{double}) is tunable: as the magnetic field strength increases, both the height of the potential barrier and the physical separation between the stable vortex states decrease. $\hat{U}_{DW}(\tau_0)$ is plotted in Figure~\ref{doublewell}(a) for $\bar{\nu}=0.55$, $c/\lambda_J=2.25$ and few negative values of the $\wp h_{\bot}$ product. For small fields, $|h_{\bot}|<1$, the field-induced shift of the potential minima can be linearized as $\tau_m \approx \pm \pi/2 (1+\wp h_{\bot} \Delta \lambda_J \cosh\bar{\nu} \sqrt{2\cosh2\bar{\nu}}/4c)$. %[[see potential.nb]] 
For large fields, eventually the minima coalesce and $\hat{U}_{DW}$ becomes single-welled. The field at which the barrier disappears completely is found by calculating when the curvature of the potential becomes positive in the origin. In Figure~\ref{doublewell}(b) we plot the field dependence of the $\hat{U}_{DW}$ in $0$, $\pm \tau_m$ and $\pm \pi$: we observe that $\hat{U}_{DW}(0)$ and $\hat{U}_{DW}(\pm \pi)$ (except for a weak quadratic term) change linearly with the perpendicular magnetic field, however, their difference is proportional to the field amplitude. For (small) negative fields the left $|L\rangle$ and right $|R\rangle$ wells of the potential constitute stable classical states for the vortex with degenerate ground state energy. For an anti-fluxon, with $\wp=-1$, we have to reverse the sign of $h_{\bot}$. Treating the fluxon as a massive classical particle, motion can only occur where its energy exceeds the potential energy; otherwise, it is trapped in one of the potential wells until the Lorentz force associated with the bias current is strong enough to start its motion. The smallest tilting that allows the vortex to escape from a well defines the so-called depinning current, $\gamma_d$. In most cases, as the vortex gets depinned, a voltage jump from zero to a finite voltage is detected. However, it may also occur that a depinned fluxon has gained a too small kinetic energy and gets trapped in the other potential well. Carapella \textit{et al.} \cite{carapella04} pointed out that the properties of a periodic, originally symmetric, tilted double-well potential are determined by the inter-well and intra-well potentials; more specifically, as far as $\hat{U}_{DW}(\pm \pi)\lesssim 2$-$3\, \hat{U}_{DW}(0)$, the $|L\rangle$ and $|R\rangle$ depinning currents are distinct, i.e., the switching to the running state occurs for different threshold currents. On the contrary, if $\hat{U}_{DW}(\pm \pi)>> \hat{U}_{DW}(0)$, it is not possible to discriminate between the two states by a current switch measurement. In our context, from Figure~\ref{doublewell}(b) we recognize that, as $h_{\bot}$ increases, we pass from a region where $\hat{U}_{DW}(\pm \pi)$ is larger than, but comparable to, $\hat{U}_{DW}(0)$, that is safe for the vortex qubit determination, to a domain, with $\hat{U}_{DW}(\pm \pi)$ much larger than $\hat{U}_{DW}(0)$, that is optimal for the qubit preparation. A detailed description of the procedures of determination and preparation of the vortex state will be given in the next Section. 

\subsection{Tunneling in the double-well potential}

\noindent In the quantum regime vortex tunneling is expected between the $|L\rangle$ and $|R\rangle$ energy levels. Since the tunneling in a double-well potential is governed by just the form of the potential in between the minima \cite{dennison32}, we are allowed to replace the profile in Eq.(\ref{double}) with a simpler expression. When the inter-well barrier, $\hat{U}_{DW}(0)$, is much larger than the intra-well height, $\hat{U}_{DW}(\pm \pi)$, the finite-wall potential in Eq.(\ref{double}) can be conveniently approximate by the infinite-wall double-well potential:
\vskip -8pt
\begin{equation}
\hat{V}_{DW}(\tau_0)=\frac{\hat{\omega_m}^2}{8\tau_m^2}(\tau_0-\tau_m)^2 (\tau_0+\tau_m)^2=\frac{\hat{\omega_m}^2}{8\tau_m^2}(\tau_0^2-\tau_m^2)^2, 
% + \textrm{const}
\label{V}
\end{equation}
\vskip -4pt
%[[see Notebooks/DoubleWellPotential.nb]]
\begin{figure}[tb]
\centering
%\subfigure[ ]{\includegraphics[height=5cm,width=7cm]{DWP.png}}
%\subfigure[ ]{\includegraphics[height=5cm,width=7cm]{omegasquared.png}}
\includegraphics[width=8cm]{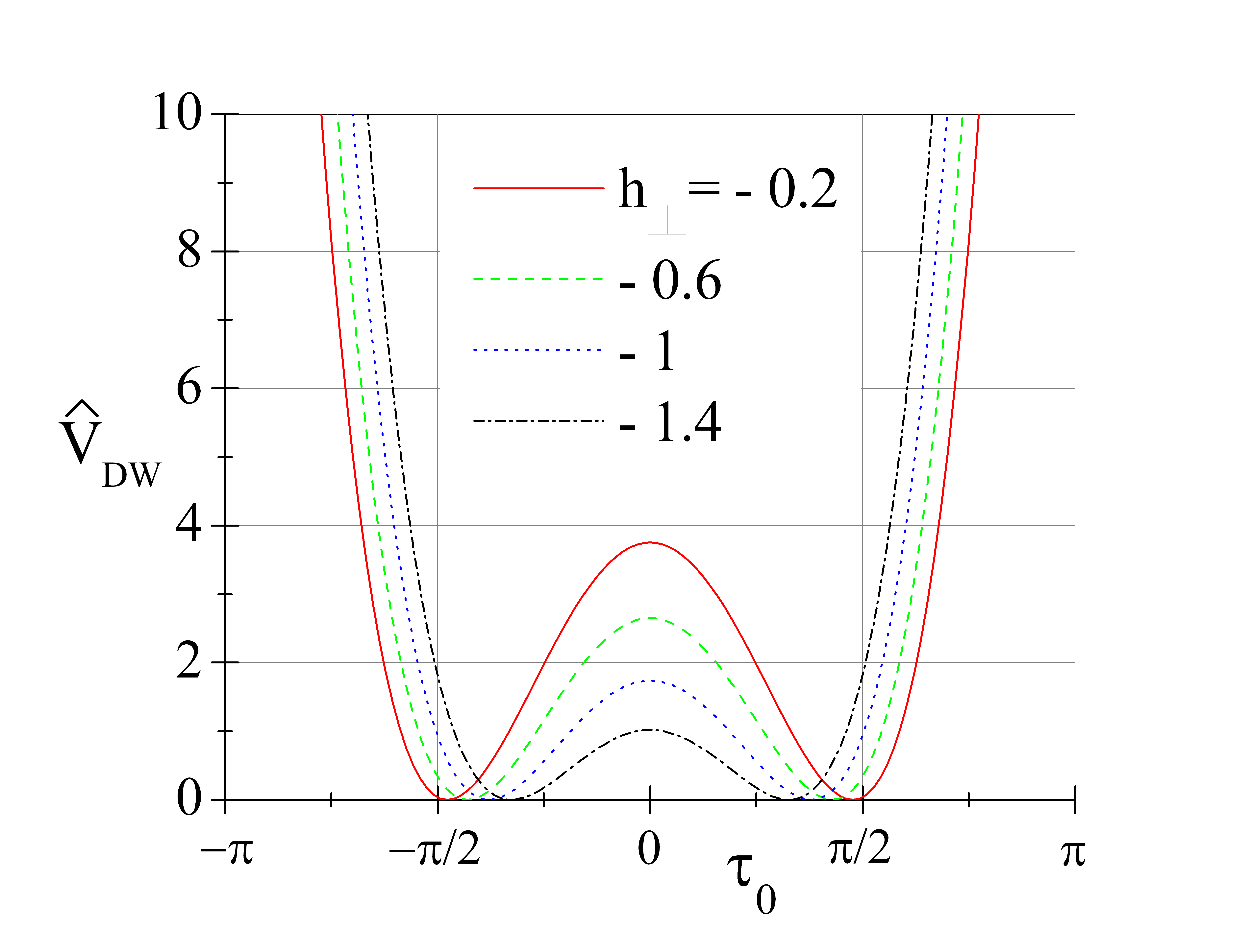}
\caption{(Color online) The infinite-wall double-well potential as in Eq.(\ref{V}) with the same parameters used for Figure~\ref{doublewell}(a).}
\label{DWP}
\end{figure}
%[[Furthermore, the field dependence of the squared normalized angular frequency $\hat{\omega_m}^2$ and of the coupling strength $g=1/\hat{\omega_m}^2 \tau_m^2=1/8\hat{U}_{DW}(0)$ are shown in Figure~\ref{doublewell}(b).]] 
\noindent which exhibits the two degenerate symmetric minima at $\tau_0 =\pm \tau_m$. Near the minima, the potential looks approximately like a shifted harmonic oscillator potential, $\hat{\omega_m}^2 (\tau_0 \pm \tau_m)^2/2$. The height of the potential barrier at the center is $\hat{V}_{DW}(0)=\hat{\omega_m}^2 \tau_m^2/8$; therefore, it must be $\hat{\omega_m}^2= 8\hat{U}_{DW}(0)/\tau_m^2$. The infinite-wall potential $\hat{V}_{DW}(\tau_0)$ is plotted in Figure~\ref{DWP}(a) using the same parameters of Figure~\ref{doublewell}(a). We observe that $\hat{V}_{DW}$ very closely reproduces $\hat{U}_{DW}$ (see Figure~\ref{doublewell}(a)) as far as $|\tau_0| \lesssim 2$. When the barrier is high compared to the energy of the vortex, then there are two degenerate states, corresponding to the particle being localized in one or the other of the wells. In the quantum limit, i.e., substantially below the crossover temperature, the two spatially distinct fluxon states may be employed as a degenerate two-state system. In the limit of small energies, $\hat{E}<< \hat{V}_{DW}(0)$, the barrier height becomes infinite and the system decomposes into a sum of two independent harmonic-oscillator potentials widely separated from each other. Correspondingly, regarding our two-state system as totally isolated from its environment, the wave functions of the system should tend to two separate sets of oscillator wave-functions, called qubit basis, $\psi_n^{|L\rangle}(\tau_0)$ and $\psi_n^{|R\rangle}(\tau_0)= \psi_n^{|L\rangle}(-\tau_0)$, each with eigenenergies $E_n=(n+1/2)\hbar \omega_m$. If the fluxon energy is smaller than, but comparable to, the inter-well barrier, a fluxon in either of the two oscillator wells has a non-vanishing amplitude for tunneling through the barrier to the other well, and the wave functions of the right- and left-hand oscillators are mixed with each other. Being the potential symmetric with respect to inversion ($\tau_0 \to -\tau_0$), the wave-functions will be divided in two classes, those which are even function of $\tau_0$ and those which are odd functions. Then the proper wave-functions are the positive and negative linear combinations of the wave-functions corresponding to the two detached wells, $\Psi_n^\pm=( \psi_n^{|L\rangle} \pm \psi_n^{|R\rangle})/\sqrt{2}$ with equal probability to find the fluxon in the left or right well. Putting the two wells into ``communication'' the degeneracy is lifted \cite{landau}: in fact, the symmetric states ($\Psi_n^+$) will have a slightly lower energy than the corresponding antisymmetric states ($\Psi_n^-$). This shows that for each level of the one minimum problem there is a pair of levels for the double minimum case. The difference in the energy level, $\Delta_n$, between the even and odd states, makes the probability $\mathcal{P}(\tau,t)$ of finding the vortex at any position $\tau$ at time $t$, to oscillate with a period $h/\Delta_n$; then, the time it takes the fluxon to tunnel from one of the wells to the other is $h/2\Delta_n$. The magnitude of the splitting of the levels depends only upon the potential curve between the two minima and increases exponentially with decreasing barrier height (that is increasing magnetic field); therefore the tunneling occurs faster when the intra-well coupling is strengthened. It is also $\Delta_n<<E_{n+1}-E_n=\hbar \omega_m/\pi$, indicating that the tunneling time between two wells is much greater that $\pi^2 (c/c) \sinh^2\bar{\nu}$, typically of the order of tens of picoseconds (information cannot be transmitted faster than the speed of light). The tunneling probability depends only upon the potential curve in tunneling region near the origin delimited by the two classical inversion points, $\tau_+<\tau_m$ and $\tau_-=-\tau_+$, where the kinetic energy, $E-V_{DW}(\tau_{\pm})$, vanishes. The potential in Eq.(\ref{V}) is parabolic (although convex) also for $|\tau_0|<\tau_m$, namely, $\hat{V}_{DW}(\tau_0)\approx \hat{\omega_m}^2( \tau_m^2/8- \tau_0^2/4)$. For such quadratic potential hill the transmission coefficient is \cite{landau} $D=1/\{1+ \exp[V_{DW}(0)-E]/\hbar \omega_m\}$.
\vskip 4pt
A generic linear superposition of the basis states, $\Psi_n=\lambda \psi_n^{|L\rangle} +\mu \psi_n^{|R\rangle}$, where $\lambda$ and $\mu$ are complex probability amplitudes, is called pure qubit state (provided that $\lambda^2+\mu^2=1$). The potentials $\hat{U}_{DW}$ or $\hat{V}_{DW}$ addressed so far are symmetric, i.e, spatially degenerate, meaning that $\lambda^2=\mu^2=1/2$. The degeneracy can be removed by means of a small bias current or a small additional magnetic field parallel to major CAJTJ axis. In such a way it is possible to control the probability amplitudes and to realize all possible quibit operation. An advantage of qubit basis states being localized in separate wells several Josephson penetration lengths away is a very long intra-well energy relaxation time between the macroscopic quantum levels \cite{kemp10,kim11}. The coherent oscillation between the basis states, the key ingredient for the realization of a qubit, has not yet been observed for Josephson vortex qubits.
 
%[[ Plynomial approximations for the double-well potential with infinite walls:
%
%\begin{footnotesize}
%$\!\hat{U}_{DW}(\tau_0) \approx \!\left(\!\wp h_{\bot} \Delta \cosh\bar{\nu} + \frac{2 \sqrt{2}}{\pi \sqrt{\cosh2\bar{\nu}}}\frac{c}{\lambda_J} \!\right)-\left(\!\frac{\wp h_{\bot} \Delta \cosh\bar{\nu}}{2}+\frac{2 \sqrt{2}}{\pi \sqrt{\cosh2\bar{\nu}}} \frac{c}{\lambda_J}\!\right) \tau_0^2+\left(\!\frac{\wp h_{\bot} \Delta \cosh\bar{\nu}}{24} +\frac{2 \sqrt{2} }{3 \pi  \sqrt{\cosh2\bar{\nu}}}\frac{c}{\lambda_J}\!\right) \tau_0^4+O[\tau_0]^6$
%
%$\!\hat{U}_{DW}(\tau_0) \approx \left(\!\frac{\wp h_{\bot} \Delta \cosh\bar{\nu}}{24} +\frac{2 \sqrt{2} }{3 \pi  \sqrt{\cosh2\bar{\nu}}}\frac{c}{\lambda_J}\!\right) \tau_0^4-\left(\!\frac{\wp h_{\bot} \Delta \cosh\bar{\nu}}{2}+\frac{2 \sqrt{2}}{\pi \sqrt{\cosh2\bar{\nu}}} \frac{c}{\lambda_J}\!\right) \tau_0^2  + \textrm{const}$
%\end{footnotesize}
%]]

\section{The numerical simulations}
\begin{figure}[t]
\centering
%\subfigure[ ]{\includegraphics[width=8cm]{.png}}
%\subfigure[ ]{\includegraphics[width=8cm]{.png}}
\includegraphics[width=7cm]{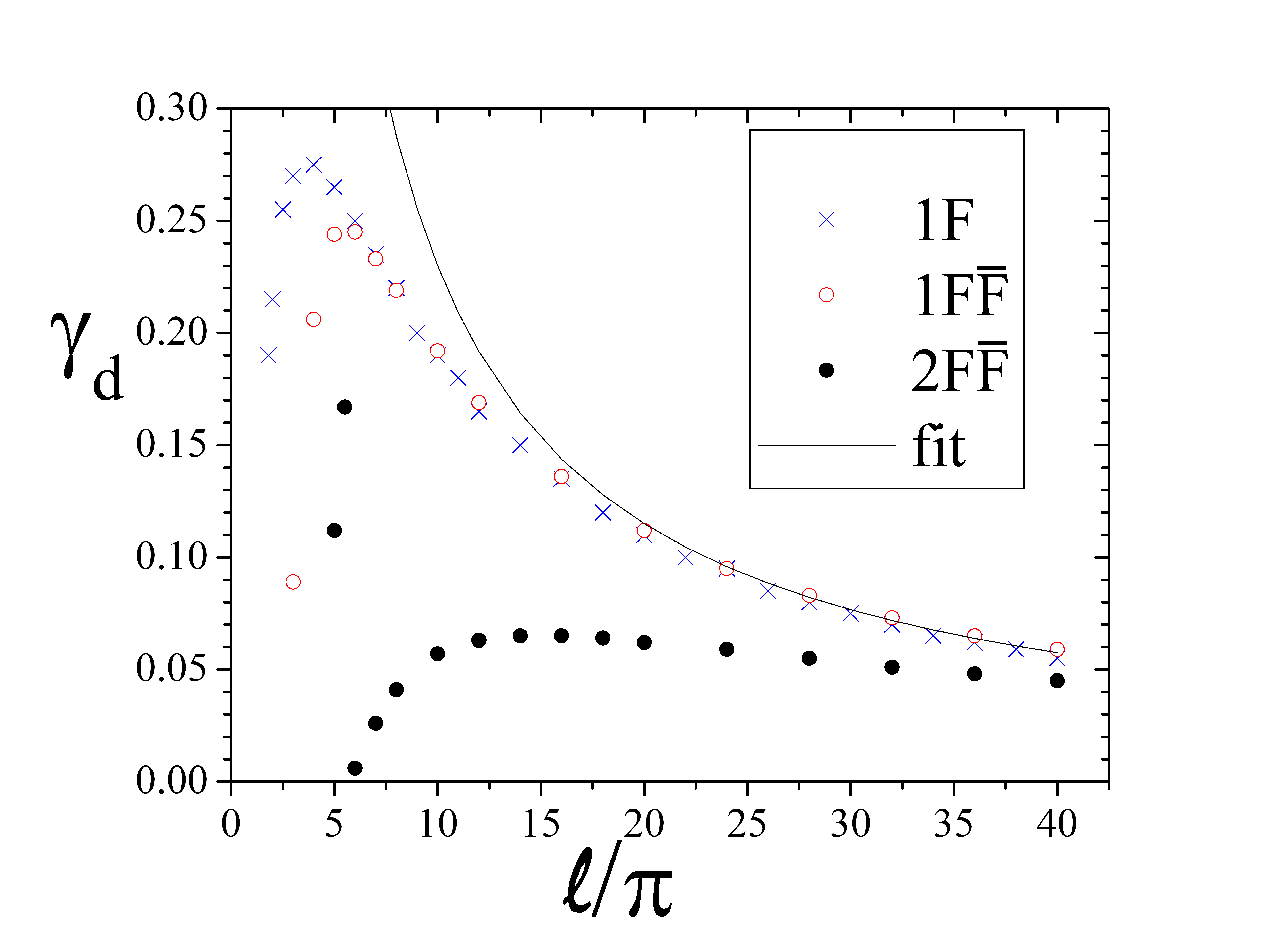}
\caption{(Color online) Numerically computed zero-field depinning currents for the static $1F$, $1F\bar{F}$ and $2F\bar{F}$ solutions versus the normalized perimeter $\ell = L/\lambda_J$ for CAJTJs having eccentricity $e^2=3/4$. The solid line is an empirical fit $\gamma_d \propto 1/\ell$ for very long junctions.}
\label{depinningH0}
\end{figure}

%Since $H_{c1}$ decreases as the annulus perimeter increases, very long CAJTJs would be unsuitable for the implementation of a robust and reliable double-well potential. 
\noindent The potential in Eq.(\ref{Utot}) has been derived under the assumption of small external field and bias current. In addition, the applied field amplitude has to be below the first critical field, $H_{c1}$, to avoid the nucleation of fluxon-antifluxon pairs that would interfere with the trapped fluxon. This implies that, for a reliable double-well operation, the longer is the normalized annulus perimeter, $\ell$, the larger must be its eccentricity. In principle, the fluxon static properties can be disclosed by minimizing the potential, i.e., by finding the roots of $d\hat{U}/d\tau_0$ and then selecting the stable $\tau_0$-positions. However, this process would only provide approximate results when $u_h$ and $u_\gamma$ cannot be considered as small perturbations and, even more, when the CAJTJ is not very long. In these cases, the potential in Eq.(\ref{Utot}) is useful just for a qualitative understanding and it is mandatory to resort to the numerical analysis. In this Section we will consider the case of a CAJTJ having an intermediate normalized length $\ell=L/\lambda_J=4\pi$, yielding $c\approx 2.25 \lambda_J$. The commercial finite element simulation package COMSOL MULTIPHYSICS (\href{url}{www.comsol.com)} was used to numerically solve Eq.(\ref{psge}) subjected to cyclic boundary conditions Eqs.(\ref{peri1}) and (\ref{peri2}) with winding number $n=1$. Although interested just in the stationary solutions, to take into account the fluxon inertial effects, we kept the $\phi_{\hat{t}}$ and $\phi_{\hat{t}\hat{t}}$ terms and run the integration for a time long enough to have $\phi_{\hat{t}}=0$. We set the damping coefficients $\alpha=0.1$ (weakly underdamped limit) and $\beta=0$, while keeping the current distribution uniform, i.e., $\gamma(\tau)= \gamma_0$. In addition, the coupling constant, $\Delta$, was set to $1$. A static fluxon centered either in $\tau_0=-\pi/2$ or $\pi/2$ was chosen for the system initial condition. Clearly, the final position of the fluxon center of mass will correspond to a (stable or meta-stable) potential minimum. In all present calculations we considered CAJTJs with eccentricity $e^2=3/4$ corresponding to $\bar{\nu}\approx 0.55$. For this choice of the parameters, the annulus width changes by a factor $2$ and the variation occurs over a length of $L/4=\pi \lambda_J$. The magnetic field dependence of the critical current of a CAJTJ has been already reported in Ref.\cite{JLTP16a} in the case of no trapped fluxons ($n=0$). At variance with any previously considered long JTJ, the zero-field critical current was found to be multiple-valued due to the existence of static fluxon-antifluxon ($F\bar{F}$) pairs with the fluxon and antifluxon in diametrically opposed points and unable to overcome the potential barriers at $\tau=0$ and $\pi$ where the annulus is widest. In Figure~\ref{depinningH0} we report the zero-field depinning currents of the $1F$, $1F\bar{F}$, and $2F\bar{F}$ solutions versus the annulus normalized perimeter. We observe that, for very long CAJTJs, the three solutions converge and the single fluxon (or anti-fluxon) depinning current decreases as $1/\ell$ (solid line). Therefore, one more reason to avoid extremely long CAJTJs is the requirement that the depinning current is an acceptable fraction of the zero-field critical current.
% Furthermore, the root-finding process (algorithm) is complex and requires numerical techniques. 

%\vskip -8pt
%\begin{equation}
%\frac{1}{2\pi}\frac{d\hat{U}}{d\tau_0}=-\frac{2\sqrt{2}}{\pi}\frac{c}{\lambda_J}\frac{\sin2\tau_0}{\sqrt{\cosh\!2\bar{\nu}+\cos\!2\tau_0}}+ h \Delta \left(\sin\bar{\theta}\sinh\bar{\nu}\cos\tau_0 -\cos\bar{\theta}\cosh\bar{\nu} \sin\tau_0 \right)+ \frac{\gamma_0}{2} (\cosh2\bar{\nu}+\cos2\tau_0)
%\label{zero}
%%\nonumber
%\end{equation}

\subsection{The vortex state determination and preparation}

\begin{figure}[t]
\centering
%\subfigure[ ]{\includegraphics[width=8cm]{.png}}
%\subfigure[ ]{\includegraphics[width=8cm]{.png}}
\includegraphics[width=7cm]{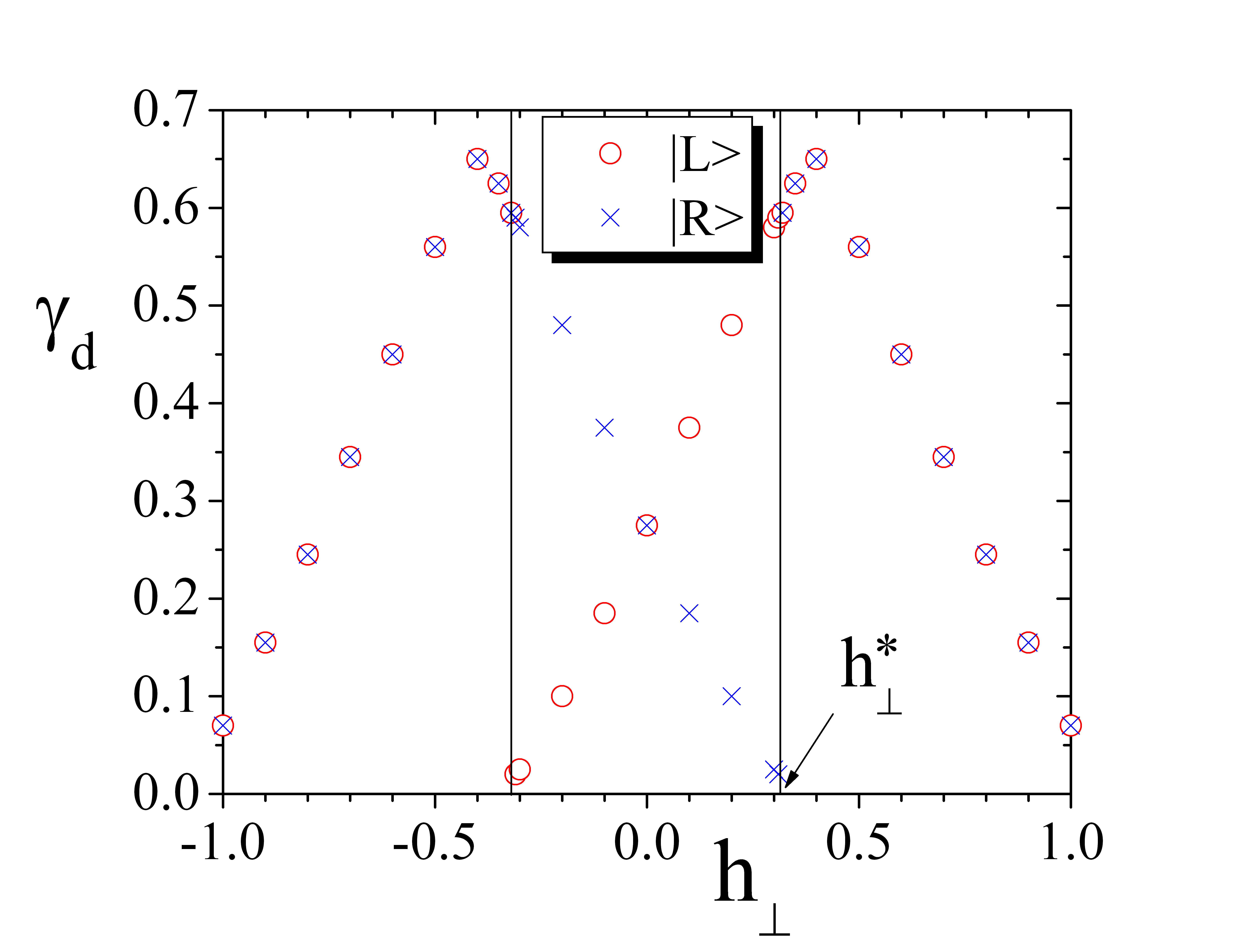}
\caption{(Color online) Numerically computed (positive) fluxon depinning currents of the $|L\rangle$ (open dots) and $|R\rangle$ (crosses) states versus the magnetic field, $h_{\bot}$, perpendicular to the longest annulus diameter.}
\label{depinningHperp}
\end{figure}

A goal of quantum information technology is to control the quantum state of a system, including its preparation, manipulation, and measurement. In this section we will investigate the state read-out and preparation processes in the presence of an in-plane field, $h_{\bot}$, applied perpendicular to the longest annulus diameter, i.e., in the direction with $\bar{\theta}=0$. We remind that the field orientation only enters in the magnetic potential $u_h$ defined in Eq.(\ref{uh}). As expected, the numerical analysis showed that, for negative $h_{\bot}$, the fluxon static positions both in the $|L\rangle$ and $|R\rangle$ states shift towards the origin until they merge for $h_{\bot}^{min}\approx -0.37$, whose absolute value is well below the (first) perpendicular critical field \cite{JLTP16b}, $|H_{c1\bot}|\approx 0.94 J_c c$; in other words, the two-minima potential safely functions in a quite wide range of operating fields, $h_{\bot}^{op}\in [h_{\bot}^{min},0]$. In Figure~\ref{depinningHperp} we report the numerically computed field dependence of the (positive) fluxon depinning currents for the $|L\rangle$ (open circles) and $|R\rangle$ (crosses) states; the positive field values are intended for an anti-fluxon. It is seen that, as far as $|h_{\bot}|<h_{\bot}^*\approx 0.31$, the fluxon escape from the $|L\rangle$ and $|R\rangle$ states occurs at quite different depinning currents, respectively, $\gamma_{d+}^{L}$ and $\gamma_{d+}^{R}$. Therefore the measure of the depinning current allows to localizes the vortex in one of the two states. Furthermore, in this range, a current inversion was found to correspond to an exchange of the $|L\rangle$ and $|R\rangle$ states, i.e., $\gamma_{d-}^{R} (h_{\bot})= -\gamma_{d+}^{L}(h_{\bot})$. It follows that the determination of the fluxon state can be as well accomplished through the measurement of a negative current switch. Figure~\ref{depinningHperp} also shows that for $|h_{\bot}|>h_{\bot}^*$ the depinning currents abruptly become identical and $\gamma_{d+}^{L}(h_{\bot}) =\gamma_{d+}^{R}(h_{\bot})= -\gamma_{d-}^{R}(h_{\bot}) =-\gamma_{d-}^{L}(h_{\bot})$. This occurs because under the action of a tilting current the fluxon escaping from one (shallow) well is trapped by the other (deep) well. Therefore, one possible procedures that can be adopted to deterministically prepare the fluxon in a given state from an arbitrary unknown initial state consists in the following four steps: i) set the amplitude of the perpendicular magnetic field (that initially was $h_{\bot}^{op}$) approximately equal to $h_{\bot}^{min}$, ii) ramp the bias current (that initially was absent) up to a value $\gamma^*$ that will be specified below; iii) restore the magnetic field to $h_{\bot}^{op}$; and iv) ramp the bias current down to zero. The polarity of $\gamma^*$ is positive (negative) for the preparation of the fluxon in the left (right) state; its amplitude has to be such that the energy acquired by the fluxon in the current-tilted potential is lower that the zero-current barrier height corresponding to the selected operating field. In such a way the fluxon remains localized in the selected state. In our simulations, with $h_{\bot}^{op}=-0.25$, we found that a preparing current $0.05\lesssim |\gamma^*|\lesssim 0.1$ reliably performed the required state operation. 

\medskip
\noindent Another possible preparing protocol involves the application of an additional magnetic field, $h_{||}$, parallel to the longest annulus diameter, i.e., with $\bar{\theta}=\pm \pi/2$ in Eq.(\ref{uh}). With $h_{\bot}$ switched off, the magnetic potential, $u_h$, is in phase with one well and out of phase with the other one; therefore, a sufficiently large field will further deepen one potential well, while removing the other one. It means that a parallel field breaks the degeneracy of the double-well potential and forces the fluxon in a given state (either $|L\rangle$ or $|R\rangle$) according to its polarity. A similar situation can also be achieved with a smaller parallel field and a properly calibrated bias current. Once the quibit state has been prepared, the parallel magnetic field (and the possible bias current) can be removed and the operating perpendicular in-plane field be restored. This alternative preparation method is less convenient since it requires the use of a second coil to generate the parallel field.

%[[The fluxon trapping current is the minimum current at which a fluxon still moves along the system, not being trapped by the potential.]]

\section{Conclusions}
\vskip -8pt

\noindent In conclusion, this paper indicates that a wide range of vortex potentials can be designed by tailoring the width of a JTL. We considered an annular JTL delimited by two closely spaced confocal ellipses that is characterized by a periodically modulated width. This configuration is faithfully modeled by a modified and perturbed one-dimensional sine-Gordon equation that, although not integrable, admits (numerically computed) solitonic solutions. The key ingredient of this partial differential equation is an effective Josephson penetration length inversely proportional to the local \jun width. This spatial dependence, in turn, generates a periodic potential that alternately attracts and repels the fluxons (or antifluxons). The potential energy minima occur at two diametrically opposite locations where the annulus is narrowest and the intra-well potential heights is uniquely determined by the CAJTJ eccentricity. When an in-plane magnetic field is applied parallel to the minor ellipse axis, a symmetric double-well potential develops with two degenerate stable states. The inter-well barrier potential height and the distance between the minima can be tuned by the field amplitude. As long as the force exerted on the vortex by the bias current is smaller than the pinning force, a pinned vortex remains confined to one of the potential minima. If the pinning force is exceeded by the driving force, the vortex starts to move. The vortex depinning current depends on the magnitude and direction of the applied field. We report on the escape of a Josephson vortex from a magnetically-tuned double-well potential. A characterization in the thermal (or classic) regime is presented here, with attention to the preparation and readout of the vortex state. We have found that when the double well potential is suitably shaped, we can prepare the vortex in a specific state by means of a particular waveform of bias current and the final state of the vortex can be read out by performing an escape measurement from one of the potential wells. Provided that the temperature and dissipation in the junction are low enough, the superposition of the macroscopically distinct states $|L\rangle$ and $|R\rangle$ can be employed to implement a reliable Josephson vortex qubit. Under sufficient decoupling from the environment, as with other superconducting qubits, the mixing between the two states, not yet observed for Josephson vortex qubits, could be identified by means of the analysis of the switching current probability distribution.

\vskip -10pt
\section*{Acknowledgments}
\noindent The author wishes to thank J. Mygind for stimulating discussions and for a critical reading of the manuscript. 
 
\newpage


\begin{thebibliography}{99}

\bibitem{hund27} F. Hund, {\it Zeits. f. Physik} {\bf 43}, 805 (1927).

\bibitem{morse} P.M. Morse, E.C.G. Stuckelberg, {\it Helvetica Physica Acta} {\bf 4}, 337 (1931); N. Rosen and P.m. Morse, {\it Phys. Rev.} {\bf 42}, 210 (1932).
%\emph{Potential fields with two minima}
%\emph{On the Vibrations of Polyatomic Molecules}.

\bibitem{dennison32} David M. Dennison and G.E. Uhlenbeck, {\it Phys. Rev.} {\bf 41}, 313 (1932).
%\emph{The Two-Minima Problem and the Ammonia Molecule}

\bibitem{Ligner} H. Lignier, C. Sias, D. Ciampini, Y. Singh, A. Zenesini, O. Morsch, and E. Arimondo, {\it Phys. Rev. Letts} {\bf 99}, 220403 (2007).

\bibitem{Theocaris} G. Theocharis, P.G. Kevrekidis, D.J. Frantzeskakis, and P. Schmelcher, {\it Phys. Rev. E} {\bf 74}, 056608 (2006).

\bibitem{Retzker} A. Retzker, R.C. Thompson, D.M. Segal, and M.B. Plenio, {\it Phys. Rev. Letts} {\bf 101}, 260504 (2008).

\bibitem{Dounas-Frazer} D.R. Dounas-Frazer, A.M. Hermundstad, L.D. Carr, {\it Phys. Rev. Letts} {\bf 99}, 200402 (2007).

\bibitem{Grossman} F. Grossmann, T. Dittrich, P. Jung, and P. Hanggi {\it Phys. Rev. Letts} {\bf 67}, 516 (1991).

\bibitem{DellaValle} G. Della Valle, M. Ornigotti, E. Cianci, V. Foglietti, P. Laporta, and S. Longhi, {\it Phys. Rev. Letts} {\bf 98}, 263601 (2007).

\bibitem{Lucero} E. Lucero,	R. Barends,	Y. Chen,	J. Kelly,	M. Mariantoni, A. Megrant,	P. O'Malley,	D. Sank, A. Vainsencher,	J. Wenner,	T. White,	Y. Yin,	A. N. Cleland	and John M. Martinis, {\it Nature Physics} {\bf 8}, 719 (2012).
%Computing prime factors with a Josephson phase qubit quantum processor

\bibitem{Makhlin} Y. Makhlin, G. Schön, A. Shnirman, {\it Nature} {\bf 398 }, 305 (1999).
%Josephson-junction qubits with controlled couplings (n.d.r charge quibit)

\bibitem{You} J.Q. You, J.S. Tsai, F. Nori, {\it Phys. Rev. Letts} {\bf 89}, 197902 (2002)
%Scalable quantum computing with Josephson charge qubits (n.d.r. theoretical)

\bibitem{Yu} Y. Yu, S.Y. Han, X. Chu, S.I. Chu, Z. Wang, {\it Science} {\bf 296}, 889 (2002).
%Coherent temporal oscillations of macroscopic quantum states in a Josephson junction (phase quibit n.d.r.)

\bibitem{Mooij} J.E.Mooij, T.P. Orlando, L. Levitov, T. Tian, C.H. van derWal, S. Lloyd, {\it Science} {\bf 285}, 1036 (1999).
%Josephson persistent-current qubit

\bibitem{kato96} Takeo Kato and Masatoshi Imada, {\it J. Phys. Soc. Jpn} {\bf 64}, 2963 (1996).

\bibitem{malomed97} A. Shnirman, E. Ben-Jacob, B. Malomed, {\it Phys. Rev. B} {\bf 56}, 14677 (1997).

\bibitem{shaju05} P.D. Shaju, V.C. Kuriakose, {\it Physica C} {\bf 424}, 125 (2005).

\bibitem{wallraff00} A. Wallraff, Yu. Koval, M. Levitchev, M.V. Fistul, and A.V. Ustinov, {\it J. Low Temp. Phys.} {\bf 118}, 543 (2000).

\bibitem{wallraff03} A. Wallraff, A. Lukashenko, J. Lisenfeld, A. Kemp, M.V. Fistul, Y. Koval and A. V. Ustinov, {\it Nature} {\bf 425}, 155 (2003).

\bibitem{kemp02}A. Kemp, A. Wallraff, A.V. Ustinov, {\it Phys. Status Solidi B} {\bf 233}, 472 (2002); A. Kemp, A.Wallraff, A.V. Ustinov, {\it Physica C} {\bf 368}, 324 (2002).

\bibitem{fistulPRB03} M.V. Fistul, and A. V. Ustinov, {\it Phys. Rev. B} {\bf 68}, 132509 (2003).

\bibitem{shaju04} P.D. Shaju, V.C. Kuriakose, {\it Phys. Letts A} {\bf 332}, 326 (2004).

\bibitem{kemp10} A.N. Price, A. Kemp, D.R. Gulevich, F.V. Kusmartsev, and A.V. Ustinov, {\it Phys. Rev. B} {\bf 81}, 014506 (2010).
%Vortex qubit based on an annular Josephson junction containing a microshort

\bibitem{kim11} Ju H. Kim and Ramesh P. Dhungana, {\it Phys. Rev. B} {\bf 83}, 064503 (2011).

\bibitem{scott} D. W. McLaughlin and A. C. Scott, {\it Phys. Rev. A} {\bf 18}, 1652 (1978).

\bibitem{Swihart} J.C. Swihart, {\it J. Appl. Phys.} {\bf 21}, 461-469 (1961).

\bibitem{nappipagano} C. Nappi and S. Pagano in \textit{National Workshop on Nonlinear Dynamics}, edited by M. Costato, A. Degasperis and M. Milani, p.261 (Italian Physical Society, Bologna, 1995).

\bibitem{landau} L.D. Landau amd E.M. Lifshitz, {\em Quantum Mechanics - Non relativistic theory}, (Pergamon Press, Oxford, 1965). 

\bibitem{gronbech} N. Gr\"{o}nbech-Jensen, P. S. Lomdahl, and M. R. Samuelsen, {\it Phys. Lett. A} {\bf 154}, 14 (1991); N. Gr\"{o}nbech-Jensen, P. S. Lomdahl, and M. R. Samuelsen, {\it Phys. Rev. B} {\bf 43}, 12799 (1991).

\bibitem{ustinov} A.V. Ustinov, {\it JETP Letters} {\bf 64}, 191 (1996).

\bibitem{PRB98} N. Martucciello, J. Mygind, V.P. Koshelets, A.V. Shchukin, L.V. Filippenko and R. Monaco, {\it Phys. Rev. B} {\bf 57}, 5444 (1998).

\bibitem{SUST15} R. Monaco, C. Granata, A. Vettoliere, and J. Mygind, {\it Supercond. Sci. Technol.} {\bf 28}, 085010 (2015).

\bibitem{JLTP16a} R. Monaco, and J. Mygind, {\it J. Low Temp. Phys.} {\bf 183}, 1 (2016). 

\bibitem{JLTP16b} R. Monaco, {\it J. Low Temp. Phys.}, DOI: 10.1007/s10909-016-1606-9; published online:\\ \url{http://link.springer.com/article/10.1007/s10909-016-1606-9}

\bibitem{davidson} A. Davidson, B. Dueholm, B. Kryger, and N. F. Pedersen, {\it Phys. Rev. Lett.} {\bf 55}, 2059 (1985).

\bibitem{dueholm} A. Davidson, B. Dueholm, and N. F. Pedersen, {\it J. Appl. Phys.} {\bf 60,} 1447 (1986).

\bibitem{hue} A. V. Ustinov, T. Doderer, R. P. Huebener, N. F. Pedersen, B. Mayer, and V. A. Oboznov, {\it Phys. Rev. Lett.} {\bf 69}, 1815 (1992).

\bibitem{PRB96} N. Martucciello, and R. Monaco, {\it Phys. Rev. B} {\bf 53} 3471 (1996).

\bibitem{sakai} S. Sakai, M.R. Samuelsen and O.H. Olsen, {\it Phys. Rev. B} {\bf 36}, 217 (1987).

\bibitem{petras} M.F. Petras and J.E. Nordman, {\it Phys. Rev. B} {\bf 39}, 6492 (1989).

\bibitem{semerdzhieva} E.G. Semerdzhieva, T. L. Boyadzhiev and Yu.M. Shukrinov, {\it J. Low Temp. Phys.} {\bf 31}, 299 (2005).
%Coordinate transformation in the model of long Josephson junctions: geometrically equivalent Josephson junctions

\bibitem{goldobin01} E. Goldobin, A. Sterck, D. Koelle, {\it Phys. Rev. E} {\bf 63}, 031111 (2001).

\bibitem{note1} It is $\hat{\mathcal{H}}(\tau,\hat{t}, \phi,\phi_\tau, \phi_{\hat{t}})d\tau = \hat{\mathcal{H}}'(\hat{s}, \hat{t}, \phi,\phi_{\hat{s}}, \phi_{\hat{t}})d\hat{s}$ where $\hat{\mathcal{H}}' (\hat{s},\hat{t},\phi,\phi_{\hat{s}}, \phi_{\hat{t}})=$\\ $=\frac{\Delta w(\hat{s})}{\lambda_J \Delta \nu}\left[ \frac{1}{2} \left( \phi_{\hat{s}} + \frac{H_\nu \Delta}{J_c c} + u_\gamma \right)^2 + \frac{1}{2} \phi^2_{\hat{t}}+1-\cos\phi \right]$ is the Hamiltonian density of Eq.(\ref{goldobin2}).

\bibitem{note2} Some useful identities for the approximate single-soliton solution in Eq.(\ref{tilde}) are reported: 

$\sin \tilde{\phi}/2 = \sqrt{(1-\cos\tilde{\phi})/2}=\sech\left[\wp (s-s_0)/\lambda_J\right]$;

$\cos \tilde{\phi}/2 = -\tanh\left[\wp (s-s_0)/\lambda_J\right]$;

$\sin \tilde{\phi} =2 \sin \tilde{\phi}/2 \, \cos \tilde{\phi}/2 =-2 \sech\left[\wp (s-s_0)/\lambda_J\right]\tanh\left[\wp (s-s_0)/\lambda_J\right]$;

%$\cos \tilde{\phi} =1-2\sin^2 \tilde{\phi}/2$;

$\tilde{\phi}_{\hat{t}}= -2\wp (\dot{s}_0/\lambda_J)\sin \tilde{\phi}/2$;

%$\tilde{\phi}_{\hat{t}\hat{t}}= -2\wp \left[(\ddot{s}_0/\lambda_J)-\wp (\dot{s}_0/\lambda_J)^2 \cos \tilde{\phi}/2 \right] \sin \tilde{\phi}/2= 2 \left[ (\dot{s}_0/\lambda_J)^2 \cos \tilde{\phi}/2 -\wp (\ddot{s}_0/\lambda_J)\right] \sin \tilde{\phi}/2$;

$\tilde{\phi}_{\hat{t}\hat{t}}= 2 \left[ (\dot{s}_0/\lambda_J)^2 \cos \tilde{\phi}/2 -\wp (\ddot{s}_0/\lambda_J)\right] \sin \tilde{\phi}/2=(\dot{s}_0/\lambda_J)^2 \sin \tilde{\phi} - 2 \wp (\ddot{s}_0/\lambda_J) \sin \tilde{\phi}/2$;

$\tilde{\phi}_\tau=-(c\mathcal{Q} /\dot{s}_0) \tilde{\phi}_{\hat{t}} =2\wp (c\mathcal{Q} /\lambda_J)  \sin \tilde{\phi}/2$;

$\tilde{\phi}_{\tau \tau}= (c \mathcal{Q}/\lambda_J)^2 \sin \tilde{\phi} - \wp (c/\lambda_J \mathcal{Q}) \sin \tilde{\phi}/2$;

$(\lambda_J/c \mathcal{Q})^2 \tilde{\phi}_{\tau\tau}-\tilde{\phi}_{\hat{t}\hat{t}}-\sin \tilde{\phi}\!=$\\$=\!\left\{2 (\dot{s}_0/\lambda_J)^2 \tanh\!\left[\wp (s\!-\!s_0)/\lambda_J\right]\!+\!2\wp (\ddot{s}_0/\lambda_J) \!-\!(\lambda_J/c\,\mathcal{Q}^3)\sin\!2\tau \right\}\sech\!\left[\wp(s\!-\!s_0)/\lambda_J\right]$.

\bibitem{benabdallah96} A. Benabdallah, J.G. Caputo, and A.C. Scott, {\it Phys. Rev. B} {\bf 54}, 16139 (1996).

\bibitem{devoret85} M.H. Devoret, J.M. Martinis, and J. Clarke, {\it Phys. Rev. Letts.} {\bf 55}, 1908 (1985).

\bibitem{cirillo16} J.A. Blackburn, M. Cirillo, N. Gr\"{o}nbech-Jensen, {\it Phys. Reps.} {\bf 611}, 1 (2016).

\bibitem{fujii08} T. Fujii, M. Nishida, and N. Hatakenaka, {\it Phys. Rev. B} {\bf 77}, 024505 (2008).

%\bibitem{koval} Y. Koval, A. Wallraff, M. Fistul, N. Thyssen, H. Kohlstedt, A.V. Ustinov, {\it IEEE Trans. on Appl. Superc.} {\bf 9}, 3957 (1999).

%\bibitem{twomicroresitors} http://meetings.aps.org/link/BAPS.2008.MAR.X10.8

%\bibitem{note3} WRONG!! For a (positive) bias current, $I$, uniformly distributed over the \jun area, $\Delta A=\pi c^2 \cosh2\bar{\nu}\Delta\nu$, it is $J_Z=I/\Delta A$ and $\gamma(\tau)= \gamma_0 \cosh2\bar{\nu}/ 2\mathcal{Q}^2(\tau)\leq \gamma_0$, where $\gamma_0\equiv (1/\Delta A)\int_{\Delta A} \gamma dS= I/J_c\Delta A \leq 1$. ($dS=c^2 \mathcal{Q}^2 \Delta\nu d\tau$)

\bibitem{ustinov06} A.A. Abdumalikov, Jr., V.V. Kurin, C. Helm, A. De Col, Y. Koval, and A.V. Ustinov, {\it Phys. Rev. B} {\bf 74}, 134515 (2006).

%\bibitem{br96} N. Martucciello, and R. Monaco, {\it Phys. Rev. B} {\bf 54} 9050 (1996).

\bibitem{PRB97} N. Martucciello, and R. Monaco, {\it Phys. Rev. B} {\bf 54} 9050 (1996); N. Martucciello, C. Soriano and R. Monaco, {\it Phys. Rev. B} {\bf 55} 15157 (1997).

\bibitem{devoret84} M.H. Devoret, J.M. Martinis, D. Esteve, J. Clarke, {\it Phys. Rev. Letts} {\bf 53}, 1260 (1984). 

\bibitem{carapella04} G. Carapella, F. Russo, R. Latempa, and G. Costabile, {\it Phys. Rev. B} {\bf 70}, 092502 (2004). 
%
\bibitem{fistulPRL03} M.V. Fistul, A.Wallraff, Y. Koval, A. Lukashenko, B. A. Malomed, and A. V. Ustinov, {\it Phys. Rev. Letts.} {\bf 91}, 257004 (2003).

%-----------------------------------------------
%\bibitem{wei} M. Weihnacht, {\it Phys. Stat Sol.} {\bf 32}, K169 (1969).
%
%\bibitem{SUST13a} R. Monaco, V.P. Koshelets, A. Mukhortova, J. Mygind, {\it Supercond. Sci. Technol.} {\bf 26}, 055021 (2013).
%
%\bibitem{PRB06}  R. Monaco, M. Aaroe, J. Mygind, R.J. Rivers, and V.P. Koshelets,  {\it Phys. Rev. B} {\bf 74}, 144513 (2006).
%
%\bibitem{PRB08}  R. Monaco, M. Aaroe, J. Mygind, R.J. Rivers, and V.P. Koshelets,  {\it Phys. Rev. B} {\bf 77}, 054509 (2008).

%-----------------------------------------------
%\bibitem{brian}  B. D. Josephson, {\it Phys. Lett.} {\bf 1}, 251 (1962).

%\bibitem{ekin} R.L. Peterson and J.W. Ekin, {\it Phys. Rev. B} {\bf 42}, 8014 (1990).

%\bibitem{barone}  A. Barone and G. Patern\`o, {\em Physics and Applications of the Josephson Effect},(Wiley, New York, 1982).

%\bibitem{goldobin12} M. Knufinke, K. Ilin,M. Siegel, D. Koelle, R. Kleiner, and E. Goldobin, {\it Phys. Rev. E} {\bf 85}, 011122 (2012).

%\bibitem{barbara} P. Barbara, R. Monaco, A.V. Ustinov, {\it J. Appl. Phys.} {\bf 79}, 327 (1996).

%\bibitem{nagatsuma83} T. Nagatsuma, K. Enpuku, F. Irie and K. Yoshida, {\it J. Appl. Phys.} {\bf 54}, 3302 (1983).

%\bibitem{magnasco} M.O. Magnasco, {\it Phys. Rev. Lett.} {\bf 71}, 1477 (1993).
%
%\bibitem{carapella01} G. Carapella, {\it Phys. Rev. B} {\bf 63}, 054515 (2001); G. Carapella, G. Costabile {\it Phys. Rev. Lett.} {\bf 87}, 077002 (2001). 

%\bibitem{nappimonaco} Ciro Nappi and Roberto Monaco, {\it Physica C} {\bf 520}, 36 (2016).

%\bibitem{pagano} S. Pagano, C. Nappi, R. Cristiano, E. Esposito, L. Frunzio, L. Parlato, G.
%Peluso, G. Pepe, and U. Scotti di Uccio, in \textit{Nonlinear Superconducting Devices and High-Tc Material}, edited by R.D. Parmentier and N.F. Pedersen, 437 (World Scientific, Singapore, 1995).

%\bibitem{nagatsuma84} T. Nagatsuma, K. Enpuku, K. Yoshida, and F. Irie, {\it Jou. Appl. Phys.} {\bf 56}, 3284 (1984).

%\bibitem{PRB10} R. Monaco, J. Mygind, V.P. Koshelets, P. Dmitriev, {\it Phys. Rev. B} {\bf 81}, 054506 (2010).

%\bibitem{davidson}  A. Davidson, B. Dueholm, B. Kryger, and N. F. Pedersen,
%Phys. Rev. Lett. {\bf 55}, 2059 (1985).
%
%\bibitem{dueholm}  A. Davidson, B. Dueholm, and N. F. Pedersen, J. Appl. Phys. 
%{\bf 60,} 1447 (1986).
%
%\bibitem{hue} A. V. Ustinov, T. Doderer, R. P. Huebener, N. F. Pedersen, B. Mayer, and V. A. Oboznov, Phys. Rev. Lett. {\bf 69}, 1815 (1992).
%
%\bibitem{ustinov}  A.V. Ustinov, JETP Lett., {\bf 64 }191 (1996).
%
%\bibitem{PRB98}  N. Martucciello, J. Mygind, V.P. Koshelets, A.V.
%Shchukin, L.V. Filippenko and R. Monaco, {\it Phys. Rev. B} {\bf 57}, 5444
%(1998).
%
%\bibitem{likharev}  K.K. Likharev, {\em Dynamics of Josephson Junctions and Circuits}\space (Gordon \& Breach Science Publishers, 1984).

%\bibitem{ustinov} I.V. Vernik, S. Keil, N. Thyssen, T. Doderer, A.V. Ustinov, H. Kohlstedt, R.P. Huebener, {\it J. Apll. Phys.} {\bf 81}, 1335 (1997); A. V. Ustinov, B. A. Malomed, and N. Thyssen, {\it Phys. Lett. A} {\bf 233}, 239 (1997).

%\bibitem{welford} W.T. Welford, {\it J. Opt. Soc. Am.} {\bf 50}, 749, (1960).

%\bibitem{born} M. Born and E. Wolf, {\em Principles of optics}, (Pergamon, London, 1980).

%\bibitem{kathuria} Y.P. Kathuria, {\it J. Opt. Soc. Am. A} {\bf 2}, 852, (1985); Y.P. Kathuria, {\it IEEE Trans. Antennas Propagat.} {\bf AP-31}, 360, (1983). 

%\bibitem{JAP07} R. Monaco, M. Aaroe, J. Mygind, and V. P. Koshelets, {\it J. Appl. Phys.} {\bf 102}, 093911 (2007).

%\bibitem{JAP08} R. Monaco, M. Aaroe, J. Mygind, and V. P. Koshelets, {\it J. Appl. Phys.} {\bf 104}, 023906 (2008).

%\bibitem{JAP10} R. Monaco, {\it J. Appl. Phys.} {\bf 108}, 033906 (2010).

%\bibitem{PRB10} R. Monaco, J. Mygind, V.P. Koshelets, P. Dmitriev, {\it Phys. Rev. B} {\bf 81}, 054506 (2010).

%\bibitem{vasenko} M.R. Samuelsen and S.A. Vasenko, {\it J. Appl. Phys.} {\bf 57}, 110 (1984).

%\bibitem{yama} T. Yamashita and Y. Onodera, {\it J. Appl. Phys.} {\bf 38}, 3523 (1976).

%\bibitem{vanDuzer}  Theodore Van Duzer, Charles W. Turner, {\em Principles of Superconductive Devices and Circuits}\space (Prentice Hall- Gale, 2nd Edition, Upper Saddle River, New Jersey, 1998).

%\bibitem{lee} G.S. Lee and A.T. Barfknecht, {\it IEEE Trans. Appl. Superc.} {\bf 2}, 67 (1992).

%\bibitem{note2} For \Jos \juns fabricated with the tri-layer technique, usually $d_t>4\lambda_t$, so $d_j/d_m \simeq \coth d_b / \lambda_{b} \coth d_b /2 \lambda_{b}$; with $d_b =2 \lambda_{b}$, it is $d_j/d_m \simeq 1.36$.

%\bibitem{owen} C.S. Owen and D.J. Scalapino, {\it Phys. Rev.} {\bf 164}, 538 (1967).

%\bibitem{dettmann} F. Dettman and P.B. Weber, {\it Phys. Stat. Sol. A} {\bf 60}, 85 (1980).

%\bibitem{franz} A. Franz, A. Wallraff, and A.V. Ustinov, {\it J. Appl. Phys.} {\bf 89}, 471 (2001).
 
%\bibitem{PRB12} R. Monaco, J. Mygind, and V.P. Koshelets, {\it Phys. Rev. B} {\bf 85}, 094514 (2012).

%\bibitem{ustinov97} A.V. Ustinov, B. A. Malomed, and N. Thyssen,  {\it Phys. Lett. A} {\bf 233}, 239 (1997).

%\bibitem{ustinov98} A. V. Ustinov, {\it Physica D}{\bf 123}, 315 (1998); M. Borromeo, G. Costantini, and F. Marchesoni, {\it Phys. Rev. E}{\bf 65}, 041110 (2002).

%\bibitem{rc} I. Rosenstein and J.T. Chen, {\it Phys. Rev. Lett.} {\bf 35} 303-305 (1975).

%\bibitem{hf} A.F. Hebard and T.A. Fulton, {\it Phys. Rev. Lett.} {\bf 35} 1310-1311 (1975).

%\bibitem{miller} S.L. Miller, Kevin R. Biagi, John R. Clem, and D.K. Finnemore, {\it Phys. Rev.} {\bf 31}, 2684 (1985).

%\bibitem{PRB09} R. Monaco, M. Aaroe, J. Mygind and V.P. Koshelets, {\it Phys. Rev. B} {\bf 79}, 144521 (2009).

%\bibitem{mercereau} J. E. Mercereau and L. T. Crane, {\it Phys. Rev. Lett.} {\bf 12}, 191 (1964). 

%\bibitem{SUST12} R. Monaco, {\it Supercond. Sci. Technol.} {\bf 25}, 115011 (2012).

%\bibitem{chang79} W.H. Chang, {\it J. Appl. Phys.} {\bf 50}, 8129 (1979).

%\bibitem{footnote} Let us remind the following identities: (a) $\coth\, z/2 + \tanh \, z/2= 2\coth\,z$, (b) $\coth\,z/2-\tanh\,z/2=2\text{csch}\,z$, (c) $\coth\,z/2= \coth\,z + \text{csch}\,z$ and (d) $\tanh\,z/2= \coth\,z - \text{csch}\,z$.

%\bibitem{meyers} N.H. Meyers, {\it Proceedings of IRE} {\bf 49}, 1640 (1961).

%\bibitem{SUST13b} R. Monaco, C. Granata, R. Russo and A. Vettoliere, {\it Supercond. Sci. Technol.} {\bf 26}, 125005 (2013).

%\bibitem{APL11} R. Monaco, J. Mygind and V.P. Koshelets, {\it Appl. Phys. Lett.} {\bf 98}, 072503 (2011).

%\bibitem{snap} H. Kroger, L.N. Smith, D.W. Jillie, {\it Appl. Phys. Lett.} {\bf 39}, 280 (1981).

%\bibitem{granata07} C. Granata, A. Vettoliere, M. Russo, {\it Appl. Phys. Lett.} {\bf 91}, 122509 (2007).

%\bibitem{caputo} J.G. Caputo, N. Flytzanis and M. Devoret, {\it Phys. Rev. B} {\bf 50}, 6471-6474 (1994).

%\bibitem{JAP95} R. Monaco, G. Costabile, and N. Martucciello, {\it J. Appl. Phys.} {\bf 77}, 2073 (1995).

%\bibitem{PRB96b} N. Martucciello, and R. Monaco, {\it Phys. Rev. B} {\bf 54} 9050 (1996).

%\bibitem{SUST14} C. Granata, A. Vettoliere, R. Monaco, {\it Supercond. Sci. Technol.} {\bf 27}, 095003 (2014).

%\bibitem{joseph} B. D. Josephson, {\it Rev. Mod. Phys.} {\bf 36}, 216 (1964).

%\bibitem{ferrel}  R.A. Ferrel, and R.E. Prange, {\it Phys. Rev. Letts.}{\bf 10}, 479 (1963).

%\bibitem{meservey} R. Meservey and P. M. Tedrow, {\it J. Appl. Phys.}{\bf 40}, 2028 (1969).

%\bibitem{scott76} A.C. Scott, F.Y.F. Chu and S.A. Reible, {\it Jou. Appl. Phys.} {\bf 47}, 3272 (1976).
%
%\bibitem{kraus}  J.D. Kraus, {\em Electromagnetics}\space (McGraw-Hill, New York, 1984).

%\bibitem{dod} A. V. Ustinov, T. Doderer, B. Mayer, R. P. Huebener, I. V. Vernik, and V. A. Oboznov, Proceedings of SQUID '92.

%\bibitem{dueholm} A. Davidson, B. Dueholm, and N. F. Pedersen, J. Appl. Phys. {\bf60}, 1447 (1986).

%\bibitem{vernik} I.V. Vernik, V. A. Oboznov, and A. V. Ustinov, Phys. Lett. A {\bf 168}, 319 (1992).

%\bibitem{malomed} A.V. Ustinov, M. Cirillo, and B.A. Malomed, Phys. Rev. {\bf B47}, 8357 (1993).

%\bibitem{zant} H. S. J. van der Zant, T. P. Orlando, S. Watanabe, and S. H. Strogatz, Phys. Rev. Lett. {\bf 74}, 174 (1994).

%\bibitem{proto} B. Dueholm, A. Davidson, C. C. Tsuei, M. J. Brady, K. H. Brown, A. C. Callegari, M. M. Chen, J. H. Greiner, H. C. Jones, K. K. Kim, A. W. Kleinsasser, H. A. Notarys, G. Proto, R. H. Wang, T. Yogi,  Proceedings of LT-17, U. Eckerm, A. Schmid, W. Weber, H. Wuhl (eds), Elsevier Science Publishers B. V., (1984).

%\bibitem{matisoo} J. Matisoo, J. Appl. Phys. {\bf 40}, 1813 (1969).

%\bibitem{alex} A.V. Ustinov, Phys. Letts. A {\bf 136}, 155 (1989).

%\bibitem{pag}  S. Pagano, Ph. D. Thesis, Tech. Univ. of Denmark, report No. S42, (1987).

%\bibitem{niels} N. Gr\"{o}nbech-Jensen, B. A. Malomed and M. R. Samuelsen,
%Phys. Rev. {\bf B46}, 294 (1992).

%\bibitem{ferrel} R.A. Ferrel and R.E. Prange, Phys. Rev. Lett. {\bf 10}, 479 (1963).

%\bibitem{schwidtal} K. Schwidtal, Phys. Rev. {\bf B2}, 2526 (1970).

%\bibitem{dino} S. Pagano, B. Ruggiero, and E. Sarnelli, Phys. Rev. {\bf B43}, 5364 (1991).

%\bibitem{ambegaokar} V. Ambegaokar and A. Baratoff, Phys. Rev. Lett. {\bf 10}, 486 (1963).

%\bibitem{ander}  P. W. Anderson and J. M. Rowell, Phys. Rev. Lett.
%{\bf 10}, 230 (1963).

%\bibitem{heb}  A. F. Hebard and T. A. Fulton, Phys. Rev. Lett. {\bf 35}
%(19), 1310 (1975).

%\bibitem{mcu}  D. E. McCumber, J. Appl. Phys. {\bf 39}, 3113 (1968).

%\bibitem{bob}  R. D. Parmentier, in {\em Solitons in action}, edited by K.
%Lonngren and A. C. Scott, Academic Press, New York, p.173 (1978).

%\bibitem{ped}  N. F. Pedersen, A. Davidson, Phys. Rev. {\bf B41}, 178, %(1990).

%\bibitem{frunzio}  {\rm R. Monaco, R. Cristiano, L. Frunzio, and C. Nappi,
%J. Appl. Phys. {\bf 71}, 1888 (1992).}

%\bibitem{dave}  A. Davidson, N. F. Pedersen, and S. Pagano, Appl. Phys. Lett. {\bf 48}, 1306 (1986).

%\bibitem{eilbeck}J.C. Eilbeck, P.S. Lomdahl, O.H. Olsen, M.R. Samuelsen, J. Appl. Phys. {\bf 57}, 861 (1985).

%\bibitem{alex} A. V. Ustinov, T. Doderer, R. P. Huebener, J. Mygind, %V. A. Oboznov, and N. F. Pedersen,  Proceedings of ASC '92.

%\bibitem{oboznov} I.V. Vernik, N. Lazarides, M.P. Sorensen, %A. V. Ustinov, N.F. Pesersen, and V. A. Oboznov, %Phys. Lett. A {\bf x}, x (1994).

%\bibitem{ziv} Z. Hermon, A. Stern, E. Ben-Jacob, Phys. Rev. B {\bf B49}, 9757 (1994).

%\bibitem{elion} W. J. Elion, J. J. Wachters, L. L. Sohn, and J. E. Mooij, Phys. Rev. Lett. {\bf 71}, 2311 (1993).

%\bibitem{shinya} S. Watanabe, S. H. Strogatz, H. S. J. van der Zant, and T. P. Orlando, Phys. Rev. Lett., (1994).

%\bibitem{jaworski} 'Influence of self-fields on the flux-flow dynamics in a long Josephson junction'- M. Jaworski, {\it Superc. Sience and Technology} {\bf 17}, 327 (2004).

\end{thebibliography}
\end{document}